\documentclass[authoryear, preprint]{interact}
\immediate\write18{makeindex \jobname.nlo -s nomencl.ist -o \jobname.nls}
\usepackage{amsmath}
\usepackage{amssymb}
\usepackage{ tipa }
\usepackage[utf8]{inputenc}
\usepackage{graphicx}
\usepackage{float}
\usepackage{caption}
\usepackage{subcaption}
\usepackage{pifont}
\usepackage{natbib}
\usepackage{geometry}
\usepackage{fleqn}
\usepackage{lineno,hyperref}
\usepackage{nomencl}
\usepackage{ifthen}
\usepackage{color}
\usepackage{color,soul}
\usepackage{makecell}
\usepackage{bm}

\usepackage{graphicx}
\usepackage{pict2e}

\makeindex
\usepackage{xcolor}
\renewcommand{\nomgroup}[1]{%
    \ifthenelse{\equal{#1}{R}}{\item[\textbf{Variables}]}{%
        \ifthenelse{\equal{#1}{G}}{\item[\textit{Greek letters}]}{}
    }
}
\makenomenclature
\modulolinenumbers[5]

\begin{document}


\title{
 Space charge and ion transport in aerosol neutralization: Toward a concentration-dependent alternative to the $N_i t$ product}


\author
{
\name
{Kunal. Ghosh\textsuperscript{a,*},
\thanks{\textsuperscript{*}CONTACT Dr. Kunal Ghosh Email: K.Ghosh@leeds.ac.uk}
Gargi Sengupta\textsuperscript{b},
Rukhsar Parveen\textsuperscript{c},
Y. S. Mayya\textsuperscript{c}} 
\affil{\textsuperscript{a}School of Earth and Environment,University of Leeds, LS2 9JT, United Kingdom \\
\textsuperscript{b}Center for Atmospheric Research, University of Oulu, P.O. Box 4500, 90014 Oulu, Finland\\
\textsuperscript{c}Department of Chemical Engineering, IIT Bombay, Mumbai 400 076, India
}
}

\maketitle


\begin{abstract}
In this study, we quantify how charged particle concentration affects the neutralization rate of aerosol particles, focusing on the role of ion dynamics shaped by internal electric fields arising from net space charge. Conventional neutralizer performance is typically evaluated using the \( N_{i}t \) product, which assumes quasi-neutral conditions and neglects electric fields from small charge imbalances. We demonstrate that internal electric fields become increasingly important at high aerosol concentrations and significantly influence neutralization dynamics.
We develop a coupled ion–aerosol transport model in a two-dimensional axisymmetric geometry that includes ion generation, convection, diffusion, recombination, attachment to aerosols, and wall loss, with self-consistent electric fields obtained from the Poisson equation. Results show that even small net charges generate electric fields that enhance ion drift and accelerate neutralization, effects not captured by traditional \( N_{i}t \)-based approaches.
Using a neutralization time metric, we find that neutralization becomes slower with increasing aerosol number concentration \( N_p \), higher initial particle charge \( q_0 \), and smaller particle diameter \( d_p \) when space charge is absent. When space charge is included, the influence of \( q_0 \) and \( d_p \) diminishes, while \( N_p \) becomes the dominant factor governing neutralization behavior. Accordingly, we propose a concentration-dependent analytical expression for mean charge relaxation that captures coupled ion–aerosol transport and space charge effects. The modeling framework presented here is applicable to laboratory instruments, industrial processes, and atmospheric environments where electrostatic interactions govern aerosol behavior.

\end{abstract}


\section{Introduction}

The neutralization of aerosols is a fundamental process with critical applications in particle size classification, environmental monitoring, and industrial emissions control \citep{ADACHI1985109}. Accurate measurement and control of aerosol charge distributions are essential for reliable particle sizing, particularly in the sub-micrometer range. One of the most precise techniques for online aerosol sizing is electrical mobility analysis, typically performed using a differential mobility analyzer (DMA) \citep{knutson1975aerosol}. This method relies on converting the electrical mobility distribution of aerosols into particle size distributions, which requires accurate knowledge of the charge distribution \citep{ADACHI1985109,reischl1991relationship}.

Aerosol charging is commonly achieved using ionizers that generate ions through either electric discharge (e.g., direct or alternating current, dielectric barrier discharge) \citep{liu1975performance,cheng1997experimental,mathon2017ozone} or irradiation sources (e.g., $\alpha$, $\beta$, soft X-ray, UV) \citep{ADACHI1985109,adachi1989bipolar,lee2005bipolar,li2011aerosol}. The resulting charged aerosol populations may exhibit either unipolar or bipolar charge distributions, with the latter being more prevalent in practical applications \citep{wang1990scanning}.

In a bipolar charging environment, two competing processes occur simultaneously: charging of neutral particles and neutralization of charged particles by oppositely charged ions. Neutralization typically dominates, driving the system toward a stationary charge distribution with a near-zero mean under ideal conditions \citep{alonso1997bipolar,kimoto2009aerosol,carsi2020numerical}. However, asymmetries in ion mobility or concentration can result in a small residual mean charge. Neutralizing particles below $10$~nm poses additional challenges due to their low charging efficiency, often below $5\%$ \citep{ADACHI1985109,alonso1997bipolar}, prompting the development of advanced neutralization technologies such as radioactive sources \citep{ADACHI1985109,adachi1989bipolar,lee2005bipolar,li2011aerosol} and AC corona chargers \citep{liu1975performance,cheng1997experimental,mathon2017ozone}.

Beyond these theoretical considerations, aerosol neutralization has broad practical relevance. It is essential for eliminating unwanted initial charges in laboratory-generated aerosols \citep{ghosh2017modeling}, investigating charge behavior in atmospheric electricity \citep{brattich2019measurements,miller2024evaluating}, and ensuring sizing accuracy in instruments such as DMAs \citep{knutson1975aerosol,ADACHI1985109,reischl1991relationship}. The efficiency of this process depends on particle size, charge level, number concentration, and the presence of space charge, that is, the net imbalance of positive and negative charges that gives rise to internal electric fields (further details in Section \ref{SI}).

A key metric for quantifying neutralization efficiency is the mean relaxation time, defined as the time required for the average particle charge to decay to a specific fraction (typically \( 1/e \)) of its initial value. Classical theories, based on dilute aerosol or single-particle models \citep{fuchs1963stationary,hoppel1986ion}, assume that each particle interacts with a uniform ionic environment. Under this framework, the rate of charge loss is governed by the product \( N_it \), where \( N_i \) is the ion concentration and \( t \) is the residence time. Using this approach, \citet{liu1986aerosol} estimated neutralization times in the continuum regime, and \citet{mayya1996variation} extended this work across a wide particle size range using size-resolved coefficients. More recently, \citet{alonso2018absence} derived an exact expression for the asymptotic neutralization rate in the free molecular regime. These studies consistently show that smaller particles require longer residence times to reach equilibrium due to their limited ion capture cross-section. However, such models do not account for the complex nonlinearities that arise in high-concentration systems.

A key limitation under dense particle conditions is ion depletion. As ions attach to aerosols, their concentration drops, leading to extended neutralization times and reduced efficiency \citep{hoppel1986ion}. This is compounded by space charge effects, which modify ion transport by generating internal electric fields. Recent work by \citet{jidenko2022effect} shows that aerosol-generated space charge can retro-control ion densities depending on the mixing geometry, offering new insights into the coupled dynamics of ion–aerosol systems.
Emerging applications in plasma processing \citep{kim2005plasma}, hot wire generators \citep{ghosh2021effect}, and electrostatic sprays \citep{tang1994structure} frequently involve highly charged aerosol flows, often exceeding the limits of conventional radioactive neutralizers. In such cases, coronal-based bipolar chargers \citep{ibarra2020bipolar} are increasingly used, yet their performance under space-charge-limited conditions is not well understood.

This study addresses these gaps by providing a detailed theoretical analysis of space charge effects on aerosol neutralization. We develop a self-consistent ion–aerosol transport model that resolves spatial and temporal variations in ion concentration and charge distributions. By explicitly incorporating internal electric fields arising from net space charge, we show that ion drift, rather than diffusion alone, governs charge relaxation in high-density systems. To isolate these effects, we compare neutralizer scenarios with and without space charge. Our simulations reveal that, particularly at high aerosol concentrations, neutralization deviates markedly from the classical single-timescale exponential decay assumed by the \( N_it \) product. Based on this, we derive a concentration-dependent analytical expression for the mean charge decay, which accounts for coupled ion–aerosol evolution and offers a more accurate and practical alternative to the \( N_it \) metric.

To our knowledge, this is the first systematic modeling framework to quantify the role of space charge in bipolar aerosol neutralization. These insights are directly relevant to the design and optimization of neutralizers for use in industrial and atmospheric settings where high particle loads and limited ion availability are prevalent.

\section{Methodology}

We developed a numerical framework to model the temporal and spatial evolution of ion and aerosol charge distributions within a cylindrical neutralizer. The system considers steady ion generation, aerosol injection, gas-phase flow, and the various transport and loss mechanisms that affect both ions and particles. The analysis aims to assess how effectively initially charged aerosols are neutralized with and without considering space charge (SC) effect in a realistic flow-through system.
\subsection{Ion transport and dynamics}

The neutralizer is modeled as a cylindrical tube of radius \( R \) and length \( L \), through which laminar airflow transports aerosol particles and ions. Ion pairs are continuously generated throughout the volume at a uniform rate \( S \) (m\(^{-3}\)s\(^{-1}\)), representative of beta-radiation sources such as Kr-85. The concentrations of positive and negative ions, denoted \( N_{i}^{+}(\mathbf{x}, t) \) and \( N_{i}^{-}(\mathbf{x}, t) \), evolve according to advection–reaction–diffusion equations that capture key physical processes: (i) convection with the background airflow, (ii) drift in response to an electric field, (iii) molecular diffusion, (iv) binary recombination of oppositely charged ions, and (v) loss to charged aerosols and the neutralizer wall.

The governing equations are:
\begin{align}
\frac{\partial N_{i}^{+}}{\partial t} + u_i \frac{\partial N_{i}^{+}}{\partial x_i} 
&\,+\, \mu^+ \nabla \cdot (E N_{i}^{+}) - D^+ \nabla^2 N_{i}^{+} \notag \\
&= S - \alpha N_{i}^{+} N_{i}^{-} - N_{i}^{+} \sum_q N_{p}(q) K^+(q) - \lambda_p N_{i}^{+}, \label{eq:positive_ions} \\[1em]
\frac{\partial N_{i}^{-}}{\partial t} + u_i \frac{\partial N_{i}^{-}}{\partial x_i} 
&\, -\, \mu^- \nabla \cdot (E N_{i}^{-}) - D^- \nabla^2 N_{i}^{-} \notag \\
&= S - \alpha N_{i}^{+} N_{i}^{-} - N_{i}^{-} \sum_q N_{p}(q) K^-(q) - \lambda_n N_{i}^{-}. \label{eq:negative_ions}
\end{align}

Here, \( u_i \) is the velocity field of the background airflow. The parameters \( \mu^\pm \) and \( D^\pm \) denote the electrical mobility and diffusion coefficients of the ions, respectively. The source term \( S \) represents the ion pair generation rate. The coefficient \( \alpha \) governs ion–ion recombination, and \( K^\pm(q) \) are the rate coefficients for ion attachment to aerosol particles in charge state \( q \). The aerosol number concentration in charge state \( q \) is denoted by \( N_{p}(q) \). The terms \( \lambda_p p \) and \( \lambda_n n \) represent first-order wall losses, with \( \lambda_p \) and \( \lambda_n \) computed using the empirical models of \citet{ghosh2020particle,ghosh2021effect}, which account for ion diffusivity, drift, and channel geometry.

The electric field \( E \) appearing in the drift terms is not externally imposed but is computed self-consistently from the local net space charge density via Poisson’s equation. This self-consistent field captures the influence of space charge on ion motion, as detailed in the following section.

\subsection{Space charge}

The electric field \( E \) is determined from the local charge distribution using Poisson’s equation:
\begin{equation}
\nabla \cdot E = -\frac{e}{\varepsilon} \left[N_{i}^{+} - N_{i}^{-} + \sum_q q N_{p}(q) \right],
\label{eq:poisson}
\end{equation}

where, \( e \) is the elementary charge, \( \varepsilon \) is the vacuum permittivity, and \( q \) denotes the discrete charge state of aerosol particles, which can take both positive and negative integer values. The term \( \sum_q q N_{p}(q) \) represents the total aerosol charge density summed over all charge states.

Eq.~\ref{eq:poisson} defines the electric field \( E \) as a response to the local net charge density, including contributions from both ions and charged aerosol particles. This self-consistent field enters the ion transport equations through the drift terms \( \mu^\pm \nabla \cdot (E N_{i}^{+}) \) and \( \mu^\mp \nabla \cdot (E N_{i}^{
-
}) \), thereby coupling ion motion directly to space charge. Including this coupling is essential for capturing the full physics of the system. In its absence, that is, under the assumption of strict quasi-neutrality where \( N_{i}^{+} \approx N_{i}^{-} \) and aerosol contributions are neglected, the system fails to achieve complete neutralization, even in a perfectly symmetric bipolar ion environment.This underscores the essential role of space-charge–induced ion drift in enabling complete charge relaxation.
This phenomenon was previously discussed in foundational work by Hoppel and Frick \citep{hoppel1986ion,hoppel1990nonequilibrium}, although the critical role of space charge in facilitating charge relaxation was not fully emphasized. Our supplementary analysis (Section \ref{SI}) demonstrates that even a minute net aerosol charge can generate an electric field sufficient to drive a drift current of ions. This drift is essential for enabling the aerosol charge distribution to relax toward neutrality. Without the space charge field, this drift mechanism is absent, and the mean aerosol charge stabilizes at a non-zero value, even when ion mobilities and attachment rates are perfectly symmetric. Thus, an apparently negligible contribution of space charge can exert a decisive influence on the system's outcome. This contradiction, wherein a small effect governs whether complete neutralization is achieved, highlights the essential role of space-charge dynamics in resolving the limitations of classical neutralization theory (see Supplementary Information, Section~\ref{SI}).

\subsection{Aerosol charging and transport}

In addition to ion dynamics, we also model the charging and transport of aerosol particles to accurately represent the neutralization process.
Monodisperse aerosols are introduced at the neutralizer inlet with an initial charge distribution. The number concentration of particles in charge state \( q \), denoted \( N_{p}(q) \), evolves due to ion attachment through a birth–death process:

\begin{align}
\frac{\partial N_{p}(q)}{\partial t} &= N_{i}^{+} K^+(q{-}1) N_{p}(q{-}1) + N_{i}^{-} K^-(q{+}1) N_{p}(q{+}1) \notag \\
&\quad - \left[ N_{i}^{+} K^+(q) + N_{i}^{-} K^-(q) \right] N_{p}(q)
\label{eq:birth_death}
\end{align}

This equation describes the charging kinetics of aerosols via ion attachment. The gain terms correspond to particles transitioning into state \( q \) from adjacent charge states, while the loss terms account for particles leaving state \( q \) due to further charging.

The spatial transport of aerosols in each charge state is governed by:

\begin{equation}
\frac{\partial N_{p}(q)}{\partial t} + \nabla \cdot (u_i N_{p}(q)) + \mu N_{p}(q) \nabla \cdot E = D \nabla^2 N_{p}(q) - \lambda_a N_{p}(q),
\label{eq:transport}
\end{equation}

where \( u_i \) is the carrier gas velocity, and \( \mu \) and \( D \) are the aerosol electrical mobility and diffusion coefficient, respectively. The term \( \lambda_a \) represents the first-order wall loss rate for aerosols. As with ion losses, \( \lambda_a \) is computed using the semi-empirical model of \citet{VOHRA2017330, ghosh2020particle,ghosh2021effect}, which accounts for both diffusive and electric field-driven migration to the wall.

\subsection{Flow field}

The carrier gas velocity field \( u_i \) is obtained by solving the incompressible Navier–Stokes equations under steady, laminar conditions:

\begin{equation}
\rho u_j \frac{\partial u_i}{\partial x_j} = -\frac{\partial P}{\partial x_i} + \mu_{\text{air}} \nabla^2 u_i,
\label{eq:momentum}
\end{equation}

where \( \rho \) is the gas density, \( P \) is the pressure, and \( \mu_{\text{air}} \) is the dynamic viscosity. In our calculations, the carrier gas is assumed to be air, and all fluid properties are defined accordingly. The resulting velocity field is used as a prescribed input in the transport equations governing ions and aerosols.

\subsection{Computational domain and numerical scheme}

The system is modeled in a two-dimensional axisymmetric cylindrical coordinate system, which efficiently captures radial and axial gradients while preserving the essential physics of flow and electrostatic interactions. This dimensionality reduces computational cost compared to full three-dimensional simulations, while still resolving key features such as boundary-layer effects and space charge distribution. The computational domain represents a cylindrical tube of fixed radius and length, as summarized in Table~\ref{table:1}. A non-uniform mesh is employed, with finer resolution near the walls to resolve steep gradients in electric field, ion concentration, and velocity.

\begin{figure}[H]
\centering
\includegraphics[width=0.95\textwidth]{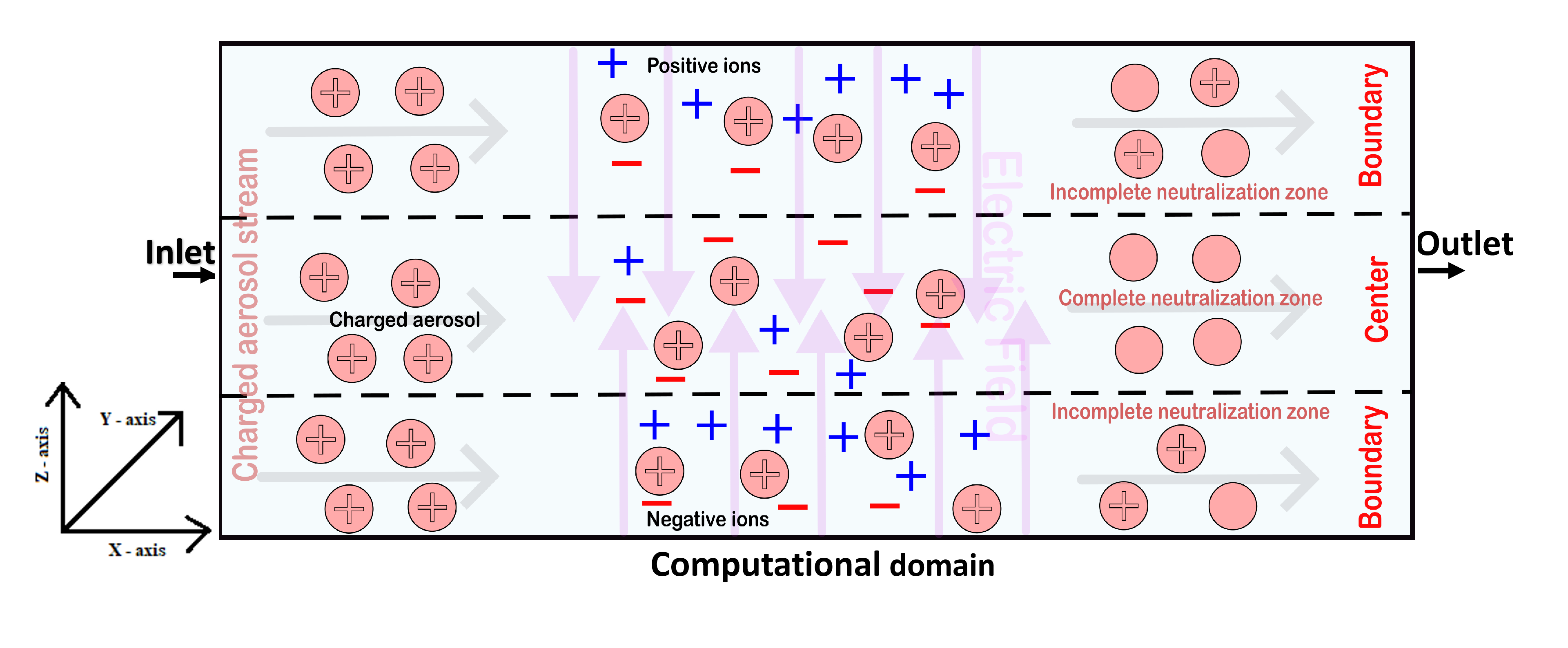}
\caption{Schematic of space-charge-assisted neutralization. Charged aerosols enter along the axis, while ions are uniformly generated throughout the volume. The electric field induced by net space charge (indicated by arrows) drives ion drift, which facilitates aerosol neutralization, especially in the central region of the flow domain.}
\label{concept}
\end{figure}

To aid visualization, a schematic of the simulation domain and key physical processes is shown in Figure~\ref{concept}. It illustrates the entry of charged aerosol particles at the inlet, uniform generation of bipolar ions throughout the volume, and the formation of an electric field from net space charge. This field induces ion drift, which enhances neutralization, particularly in the central flow region. The figure also highlights the importance of self-consistent coupling between charge distribution and electric field, a central aspect of the neutralization mechanism studied here.

To represent a range of real-world conditions, we simulate four aerosol concentrations: (a) $10^{9}$~m$^{-3}$ and (b) $10^{10}$~m$^{-3}$, representative of typical ambient atmospheric levels \citep{mcmurry2000review}; and (c) $10^{11}$~m$^{-3}$ and (d) $10^{12}$~m$^{-3}$, corresponding to concentrations observed in industrial and laboratory settings such as vehicular exhaust \citep{wehner2009aerosol}, chimney plumes \citep{chen2017stack}, and laboratory particle generators \citep{ghosh2020particle,ghosh2021effect}. In all simulations, aerosol particles are continuously injected at the inlet for the first 1~second to mimic the entry of a short aerosol pulse into the neutralizer. After this injection phase, no additional particles are introduced, and the system evolves under the influence of flow, diffusion, and electrostatic interactions.

The ion production rate is fixed at $10^{12}$ ions m$^{-3}$s$^{-1}$, and the flow rate is set to 2 liters per minute (LPM). These values are based on operating conditions of commercial bipolar neutralizers that use Kr-85 or soft X-ray sources \citep{ADACHI1985109,reischl1991relationship}. The selected flow rate provides typical residence times sufficient for ion–aerosol interaction without inducing turbulence in compact laboratory-scale setups.

\begin{table}[H]
\centering
\caption{Model parameters used in simulations}
\label{table:1}
\begin{tabular}{p{1.2in} p{2.0in} p{2.0in}}
\hline
\textbf{Item} & \textbf{Parameter Name} & \textbf{Value(s)} \\
\hline
Neutralizer & Length ($L$) & 0.1 m \\
            & Radius ($R$) & 0.03 m \\
            & Flow rate & 2 LPM \\
Ion generation & Ion production rate ($S$) & $10^{12}$ ions m$^{-3}$s$^{-1}$ \\
Aerosols & Initial concentration ($N_p$) & $10^{9}$–$10^{12}$ m$^{-3}$ \\
         & Initial charge level ($q$) & +10 elementary charges \\
         & Particle diameter ($d_p$) & 1 $\mu$m (monodisperse) \\
\hline
\end{tabular}
\end{table}

The governing equations for gas flow (Navier–Stokes), ion transport, and aerosol charging are solved sequentially at each time step using a semi-implicit finite difference scheme, which balances computational efficiency with stability for stiff, coupled equations. To ensure accuracy near boundaries, enhanced wall-resolution techniques are applied to better resolve diffusion and drift effects in regions with strong gradients. The system is assumed to operate under isothermal conditions, with constant temperature throughout the computational domain.

\section{Results}

\begin{figure}[H]
    \centering
    \includegraphics[width=\textwidth]{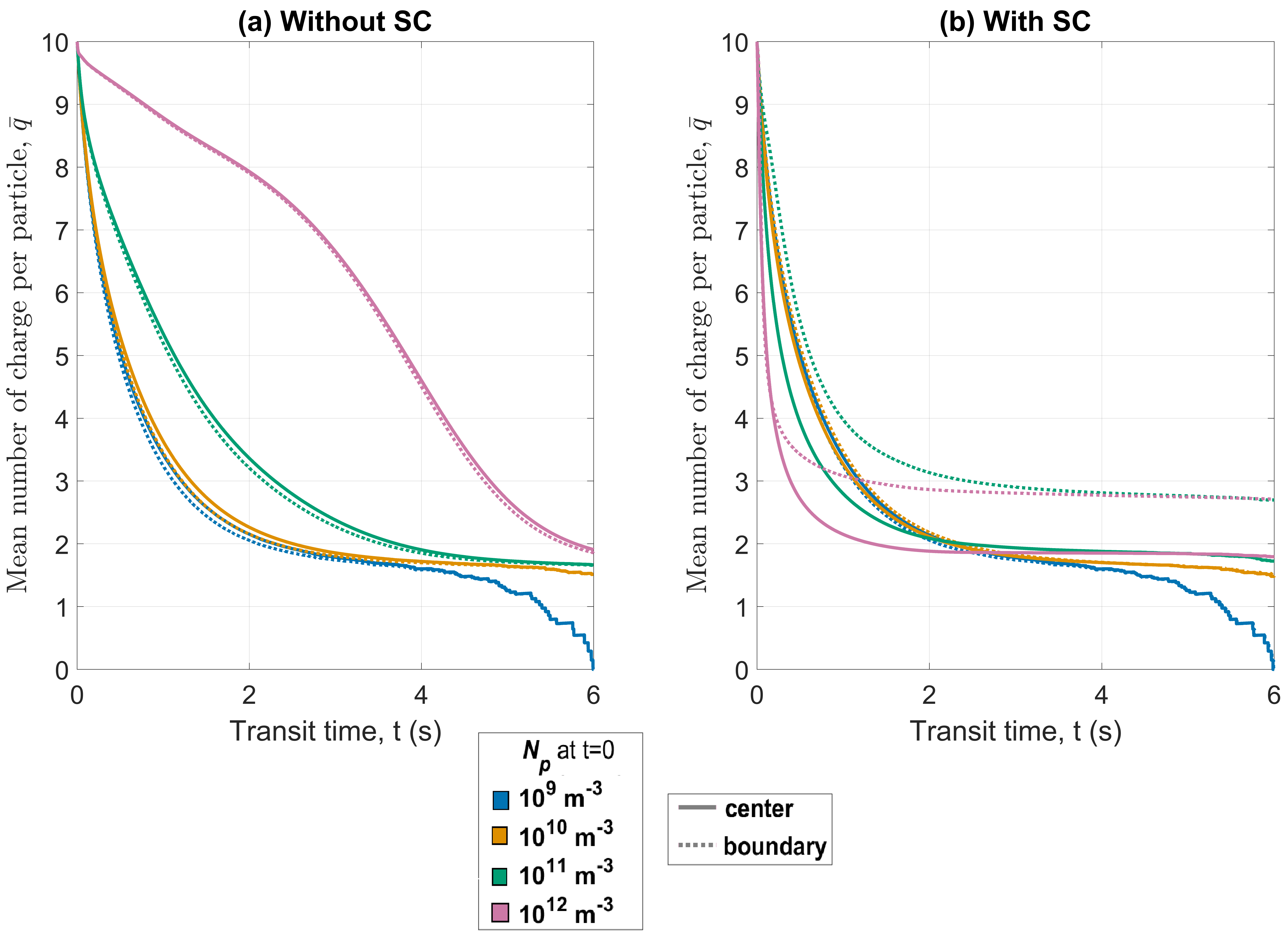}
    \caption{Time evolution of the mean number of charges per particle ($\bar q$) at the center (solid lines) and boundary (dotted lines) of the neutralizer. 
Simulations are shown for four initial aerosol number concentrations: 
\( N_p\big|_{t=0} = 10^{9}~\mathrm{m}^{-3} \) (blue), $10^{10}$~m$^{-3}$ (orange), $10^{11}$~m$^{-3}$ (green), and $10^{12}$~m$^{-3}$ (pink). 
Panel (a) shows results without space charge (SC) effects, and panel (b) includes SC effects. 
All simulations were conducted at an ion concentration of $N_i = 10^{12}~\mathrm{m}^{-3}$, assuming a monodisperse aerosol population with a particle diameter of $d_p = 1~\mu\mathrm{m}$.}
    \label{qmean}
\end{figure}

Figure~\ref{qmean} compares the temporal evolution of the mean number of charges per particle ($\bar q$) at the center and boundary of the neutralizer for four different initial aerosol concentrations (\( N_p\big|_{t=0} \)), both without (panel a) and with (panel b) space charge (SC) effects. Simulations assume a fixed ion concentration of $10^{12}$~m$^{-3}$ and monodisperse aerosols of diameter 1~$\mu$m. The 6-second duration reflects the residence time corresponding to the specified flow rate and geometry, allowing full aerosol transit through the device.

In the absence of SC (panel a), $\bar q$ decays smoothly over time, with center and boundary profiles showing similar trends across all concentrations. This behavior is consistent with classical bipolar charging theory, which assumes uniform ion distributions and negligible electric fields. At higher concentrations ($10^{11}$ and $10^{12}$~m$^{-3}$), neutralization slows considerably, leaving substantial residual charge even after 6 seconds.

With SC included (panel b), strong deviations emerge. At high aerosol concentrations, centerline particles neutralize substantially faster than those near the wall, demonstrating space-charge-assisted neutralization. This mechanism, derived analytically in the Supplementary Information (Section~\ref{SI}), is confirmed numerically here. The net space charge creates internal electric fields that drive ion drift toward the center, enhancing charge relaxation in the core while depleting ions near the boundary.
This asymmetry also explains a transient boundary effect. During early times, $\bar q$ near the wall can initially increase, a counter-neutralization effect, due to positive ions being drawn toward the periphery by the field generated by positively charged aerosols. As central particles neutralize and the field weakens, the boundary charge begins to decay.

Notably, Figure~\ref{qmean}b shows that for \( N_p = 10^{12}~\mathrm{m}^{-3} \), $\bar{q}$ drops from 10 to approximately 3 much more rapidly under SC conditions. This early-stage acceleration is driven by strong internal electric fields that enhance ion mobility and facilitate rapid ion-particle interactions. As $\bar q$ decreases, the field weakens, slowing the rate of decay—consistent with the field attenuation observed in Figure~\ref{fig:normalized_efield} (Supplementary Information). These trends indicate that transient, field-driven drift dominates early neutralization, while ion depletion limits the later stages.

The spatial variation in neutralization is further supported by the electric field profiles in Figure~\ref{fig:normalized_efield}, which show that the normalized field remains stronger and more persistent near the boundary at higher aerosol concentrations. This reflects the localized buildup of space charge and its role in modulating ion transport.

Overall, the model predicts that aerosols near the center are consistently neutralized more rapidly than those near the wall, illustrating the fundamentally non-uniform nature of charge relaxation in the presence of space charge.

\begin{figure}[H]
    \centering
    \includegraphics[width=\textwidth]{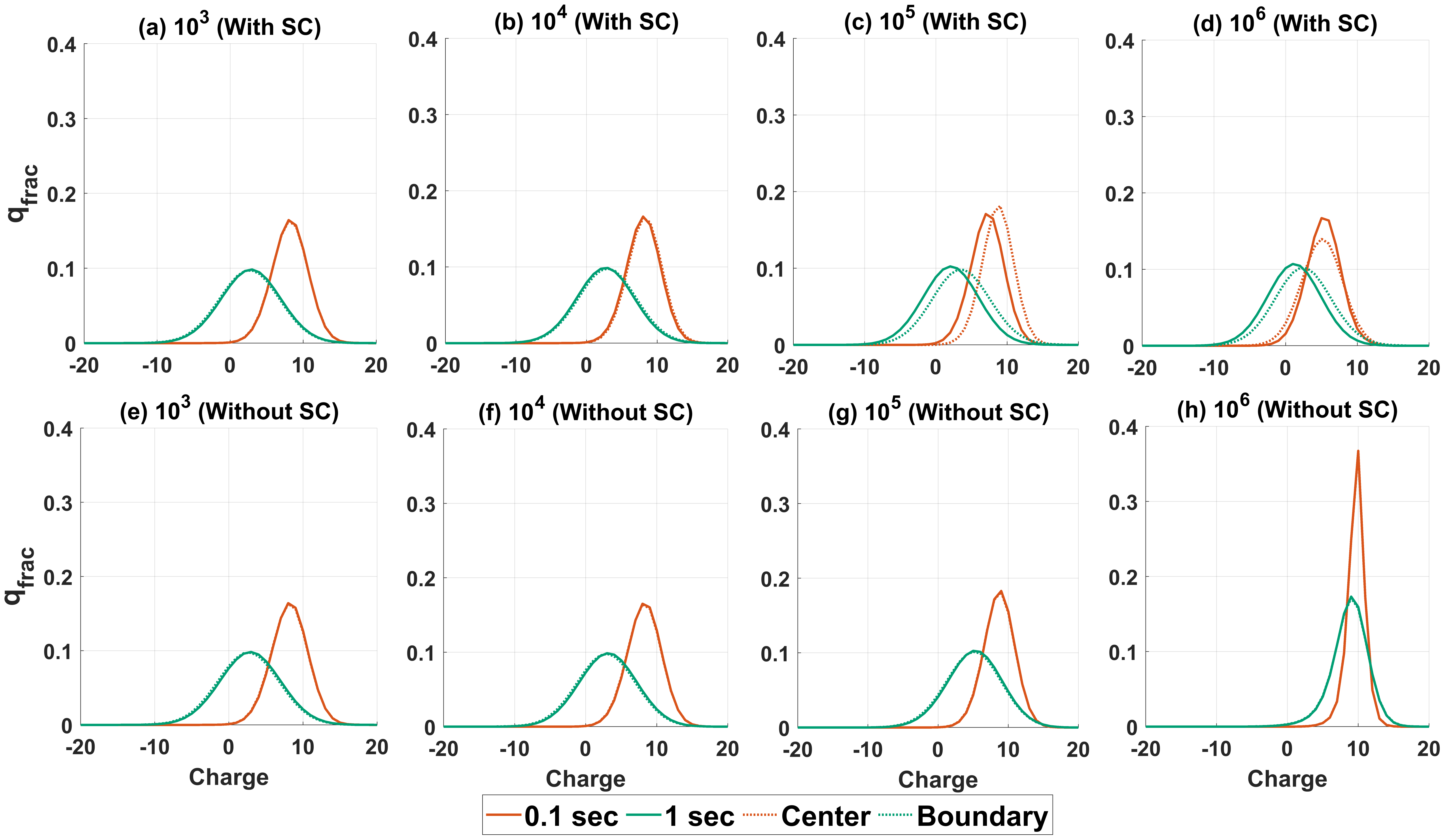}
    \caption{Charge distribution (\( q_{\text{frac}} \)) on aerosols at the center and boundary of the neutralizer at two time points (0.1~s and 1~s), for varying initial aerosol number concentrations (\( N_p\big|_{t=0} = 10^{9} \) to \( 10^{12}~\mathrm{m}^{-3} \)). Panels (a)–(d) correspond to simulations with space charge (SC) effects included, while panels (e)–(h) represent the corresponding cases without SC. All simulations were conducted at a fixed ion concentration of \( N_i = 10^{12}~\mathrm{m}^{-3} \), assuming a monodisperse aerosol population with a particle diameter of \( d_p = 1~\mu\mathrm{m} \).
}
    \label{qfrac}
\end{figure}

Figure~\ref{qfrac} shows the evolution of the aerosol charge distribution (\( q_{\text{frac}} \)) at the center and boundary of the neutralizer at two time points: 0.1~s and 1~s, across four different initial aerosol concentrations (\( N_p\big|_{t=0} = 10^{9} \) to \( 10^{12}~\mathrm{m}^{-3} \)). The discrete charge distribution \( q_{\text{frac}} \) represents the fraction of particles in the system that carry a specific integer charge \( q \), and is defined as:
\begin{equation}
q_{\text{frac}}(q) = \frac{N(q)}{N_p}
\end{equation}
where \( N(q) \) is the number of particles with charge \( q \), and \( N_p \) is the total number of particles in the system at the respective time point. The charge values \( q \) range from \(-20\) to \(+20\). Panels (a)–(d) correspond to simulations with space charge (SC) effects included, while panels (e)–(h) show the corresponding cases without SC. All cases were simulated at an ion concentration of \( N_i = 10^{12}~\mathrm{m}^{-3} \), assuming a monodisperse aerosol population with a particle diameter of \( d_p = 1~\mu\mathrm{m} \).

In the absence of SC (bottom row), the charge distributions remain spatially symmetric between the center and boundary, and their evolution follows a predictable relaxation pattern toward lower charge states. Without SC, the 1~s charge distribution does not approach a Boltzmann distribution, as classical theory might predict, but rather remains closer to a bipolar form at higher concentrations ($10^{11}$ and $10^{12}$~m$^{-3}$). At lower concentrations ($10^{9}$ and $10^{10}$~m$^{-3}$), the distributions do converge toward a Boltzmann-like shape, consistent with quasi-equilibrium conditions. This indicates that high-concentration systems do not reach full thermodynamic equilibrium within the finite residence time of the neutralizer and highlights the limitations of steady-state assumptions for fast-flowing, high-density aerosol streams.

In contrast, when SC effects are included (top row), significant spatial and temporal asymmetries emerge, especially at higher concentrations. At 0.1~s, particles near the boundary exhibit a sharply peaked, positively skewed distribution relative to centerline particles, which have already begun shifting toward neutrality. This early-stage behavior is indicative of a counter-neutralization effect: space charge–induced electric fields drive positive ions toward the periphery, increasing unipolar charging near the wall before neutralization eventually dominates. As a result, the charge fraction is initially higher at the boundary than at the center, particularly at concentrations of $10^{11}$ and $10^{12}$~m$^{-3}$.
By 1~s, centerline particles in SC cases show a broader and more neutralized distribution across all concentrations, confirming the enhanced ion accessibility and accelerated charge relaxation at the core. Meanwhile, boundary distributions remain skewed and more positively charged, particularly at high concentrations, due to localized ion depletion and field-driven ion redistribution.

The contrast between the SC and no-SC cases becomes especially stark at $10^{12}$~m$^{-3}$ [panels (d) and (h)], where the SC-driven dynamics result in strongly nonuniform charge profiles, while the no-SC case produces a sharply peaked, nearly symmetric distribution at both center and boundary. The results show that space charge not only accelerates neutralization at the center, but also induces complex spatial gradients in charge state evolution.

\begin{figure}[H]
    \centering
    \includegraphics[width=\textwidth]{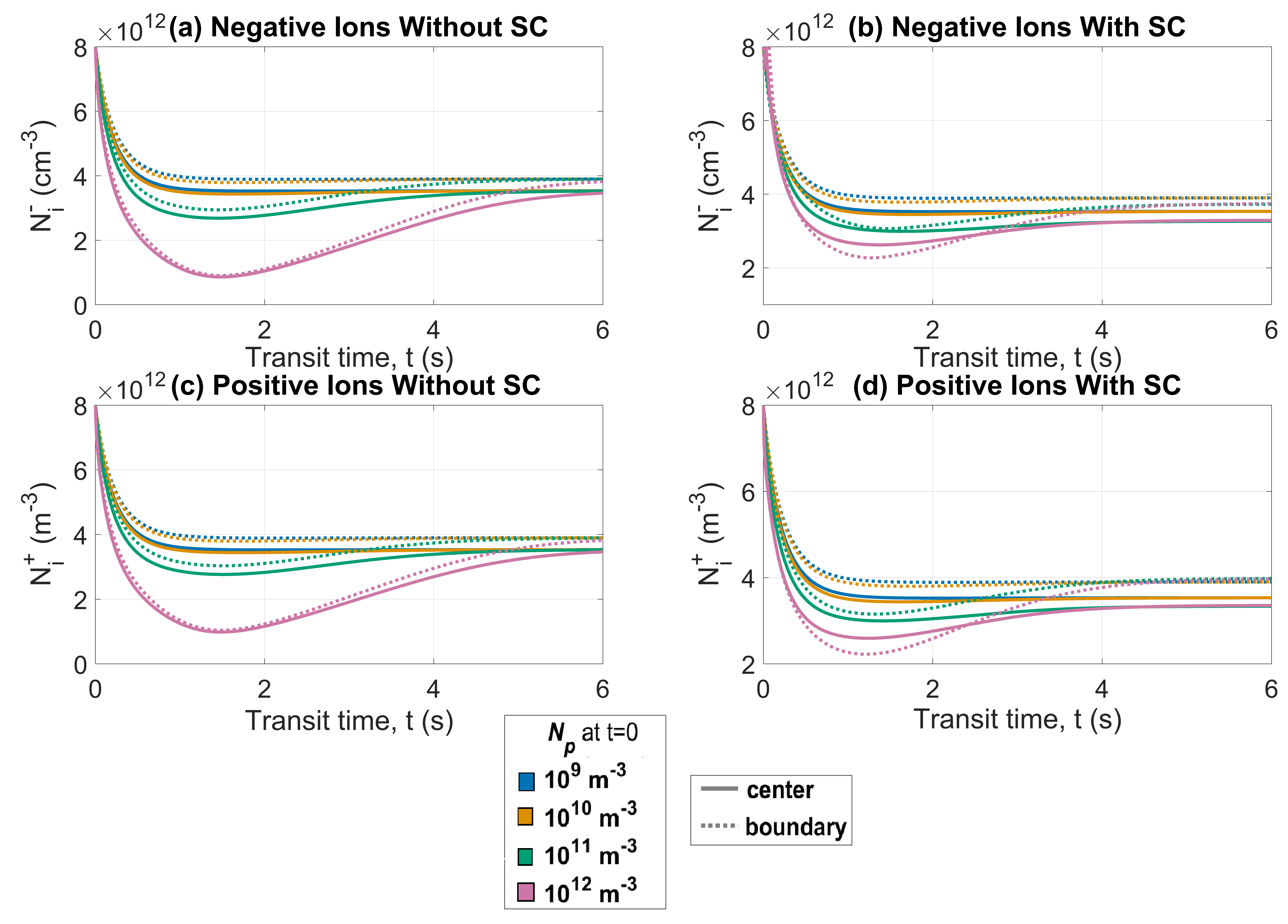}
    \caption{Time evolution of positive and negative ion concentrations ($N_i^+$ and $N_i^-$) at the center (solid lines) and boundary (dotted lines) of the neutralizer, for four initial aerosol number concentrations: 
\( N_p\big|_{t=0} = 10^{9}~\mathrm{m}^{-3} \) (blue), $10^{10}$~m$^{-3}$ (orange), $10^{11}$~m$^{-3}$ (green), and $10^{12}$~m$^{-3}$ (pink). 
Panels a and b show negative ion concentrations without and with space charge effects, respectively. 
Panels c and d show the corresponding positive ion concentrations without and with space charge effects. All simulations were conducted at an ion concentration of $N_i = 10^{12}~\mathrm{m}^{-3}$, assuming a monodisperse aerosol population with a particle diameter of $d_p = 1~\mu\mathrm{m}$.}
    \label{ionprofile}
\end{figure}

Figure~\ref{ionprofile} shows the time evolution of negative ion concentration ($N_i^-$) and positive ion concentration ($N_i^+$) at the center (solid lines) and boundary (dotted lines) of the neutralizer for varying initial aerosol number concentrations: \( N_p\big|_{t=0} = 10^{9} \) to $10^{12}$~m$^{-3}$. Panels (a) and (c) correspond to simulations without space charge (SC), while panels (b) and (d) include SC effects. For all simulations, we assumed an ion concentration of $N_i = 10^{12}~\mathrm{m}^{-3}$ and a monodisperse aerosol distribution with particle diameter $d_p = 1~\mu\mathrm{m}$.

In the absence of SC [panels (a) and (c)], both positive and negative ion concentrations decay in a spatially uniform manner. The center and boundary profiles closely overlap, indicating symmetric ion behavior as expected under classical bipolar charging conditions. The depletion is more rapid for higher aerosol concentrations due to increased ion–particle attachment. Notably, for the $10^{12}$~m$^{-3}$ case, a distinct dip is observed at early times followed by partial recovery. This reflects an initial imbalance where rapid ion attachment dominates over generation and recombination, before stabilizing toward a quasi-steady state.

When SC is included [panels (b) and (d)], strong spatial asymmetry in ion behavior emerges, particularly at high aerosol concentrations ($\geq 10^{11}$~m$^{-3}$). Ions are depleted faster near the boundary due to electric fields induced by net space charge, which drive ions toward the centerline. This SC-induced drift not only accelerates ion loss near the wall but also limits ion availability for neutralization in that region. As a result, ion concentrations at the center are maintained at higher levels relative to the boundary.

For the highest concentration case ($10^{12}$~m$^{-3}$), this asymmetry is most pronounced. The early-time dip in ion concentration is deeper at the boundary and recovery is slower compared to the center. Unlike the no-SC scenario, where the recovery is relatively uniform, the SC case shows persistent disparity between center and boundary due to the sustained action of drift-driven redistribution.

Overall, these results highlight how internal electric fields generated by space charge fundamentally alter the spatiotemporal dynamics of ion populations. The enhanced centerline retention of ions under SC leads to localized neutralization advantages, while the boundary suffers from ion starvation. This mechanism directly connects to the neutralization asymmetry observed in Figure~\ref{qmean}, and reinforces the need to incorporate space charge effects when modeling or designing aerosol neutralizers under dense particle loading conditions.

\begin{figure}[H]
    \centering
    \includegraphics[width=\textwidth]{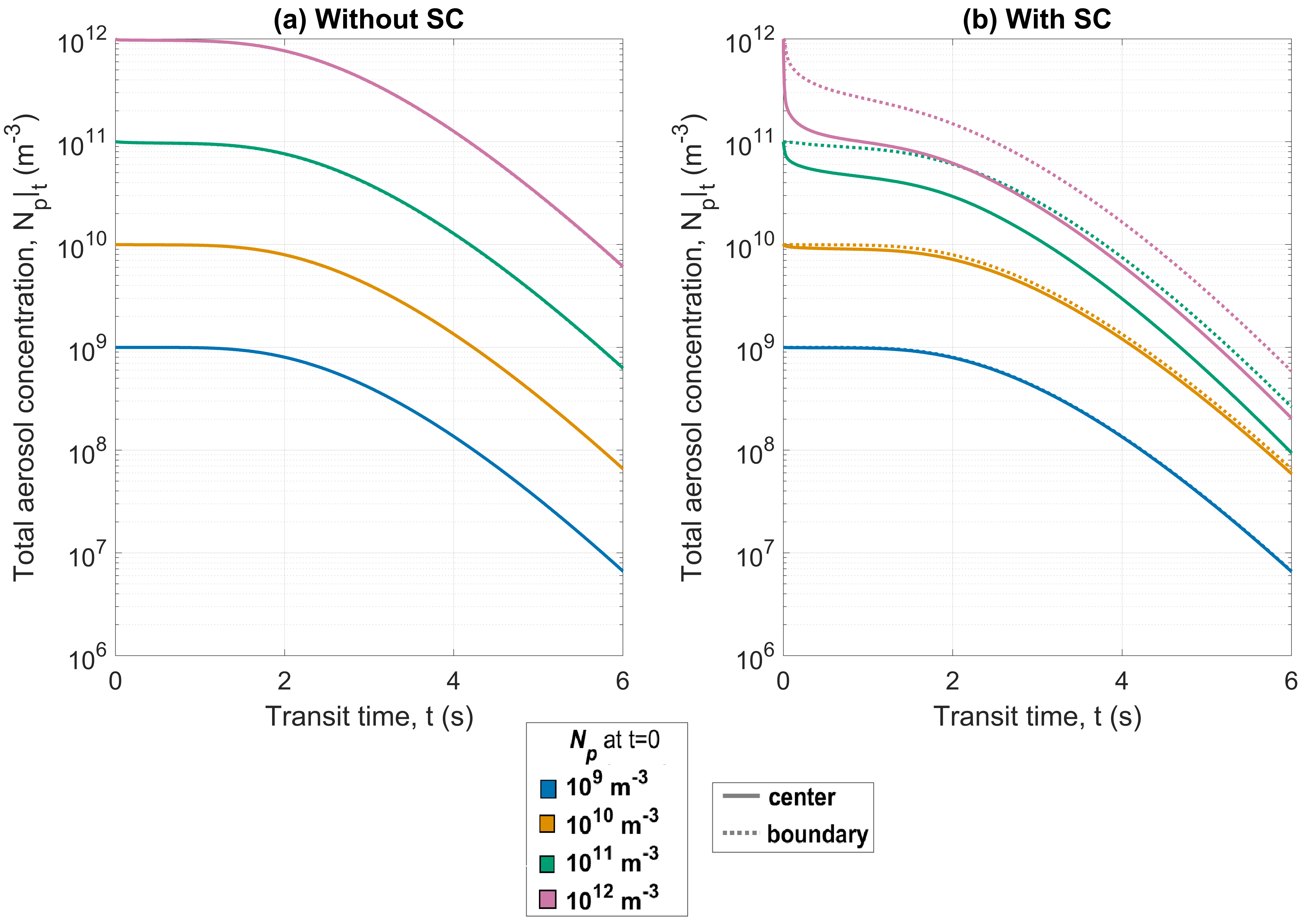}
    \caption{Time evolution of total aerosol number concentration ($N_p\big|_t$) at the center (solid lines) and boundary (dotted lines) of the neutralizer shown as a function of the transit time ($t$), for different initial aerosol concentrations: 
\( N_p\big|_{t=0} = 10^{9}~\mathrm{m}^{-3} \) (blue), $10^{10}$~m$^{-3}$ (orange), $10^{11}$~m$^{-3}$ (green), and $10^{12}$~m$^{-3}$ (pink). 
Panel a shows results without space charge effects, and panel b includes space charge effects. All simulations were conducted at an ion concentration of $N_i = 10^{12}~\mathrm{m}^{-3}$, assuming a monodisperse aerosol population with a particle diameter of $d_p = 1~\mu\mathrm{m}$.
}
    \label{aerosolconc}
\end{figure}

Figure~\ref{aerosolconc} presents the time evolution of the total aerosol number concentration (($N_p\big|_t$) as a function of the transit time ($t$), including both charged and neutral aerosols, at the center and boundary of the neutralizer for varying initial aerosol number concentrations \( N_p\big|_{t=0} = 10^{9} \) to $10^{12}$~m$^{-3}$. Simulations are shown for both without (panel a) and with (panel b) space charge (SC) effects. All cases correspond to an ion concentration of $N_i = 10^{12}~\mathrm{m}^{-3}$ and a monodisperse aerosol distribution with particle diameter $d_p = 1~\mu\mathrm{m}$.

In the absence of SC [panel (a)], aerosol concentrations decline gradually over time, driven by convection and diffusion, with negligible difference between the center and boundary profiles. The near overlap of these curves reflects spatial uniformity in aerosol transport, consistent with purely laminar flow without additional forces. The decay is smooth and monotonic across all concentrations, and higher particle loads exhibit proportionally larger losses, primarily due to the finite residence time within the neutralizer.

When SC is included [panel (b)], significant deviations emerge, especially for the high aerosol concentrations ($10^{11}$ and $10^{12}$~m$^{-3}$). In these cases, aerosol concentrations drop more sharply and diverge between the center and boundary. This behavior results from ion-driven electrostatic forces: the accumulation of space charge generates strong internal electric fields that exert outward-directed drift forces on charged aerosols. These forces push particles toward the wall, where they are lost by deposition. Consequently, particle depletion is most pronounced at the centerline, while boundary regions retain a higher aerosol concentration.

The divergence between center and boundary aerosol concentration becomes more pronounced with increasing particle loading, reflecting the nonlinear coupling between charge density, electric field strength, and aerosol migration. At the highest concentration ($10^{12}$~m$^{-3}$), central aerosol concentrations decrease by nearly two orders of magnitude over the 6-second residence time, underscoring the effectiveness of SC-induced deposition in reducing the aerosol population.

In contrast, for the lower concentrations ($10^9$ and $10^{10}$~m$^{-3}$), the influence of SC is weak or negligible. Both center and boundary profiles evolve similarly to the no-SC case, indicating that electrostatic effects are insufficient to disrupt the convective-diffusive balance. The transition between these regimes illustrates the critical role of particle loading in determining whether SC will significantly impact aerosol transport.

It is important to note that the underlying airflow field, computed from a steady-state, incompressible Navier–Stokes solution, is fully developed, symmetric, and parabolic (see Supplementary Figure~\ref{fig:velocity profile}). This velocity profile implies equal convective transport toward both the center and boundary in the absence of external forces. Therefore, the radial aerosol redistribution observed under SC cannot be attributed to fluid dynamics but is instead a direct consequence of electrostatic drift. These findings reinforce the need to account for space charge effects in high-concentration aerosol systems, particularly when assessing particle residence time, loss mechanisms, and neutralizer performance.

Notably, in Figure~\ref{aerosolconc}(b), the centerline aerosol concentration profiles for \( N_p = 10^{11} \) and \( 10^{12}~\mathrm{m}^{-3} \) overlap substantially throughout the residence time. This behavior suggests that beyond a certain particle concentration threshold, aerosol loss is primarily governed by space-charge-induced radial drift and wall deposition, rather than by convective or diffusive transport alone. The convergence of these decay curves indicates that increasing particle loading beyond \( 10^{11}~\mathrm{m}^{-3} \) does not linearly extend neutralization time, and that SC-driven drift saturates the depletion mechanism at the centerline. These results support the interpretation that neutralization dynamics under SC are strongly nonlinear and dominated by internal electrostatic fields at high \( N_p \).

\begin{figure}[H]
    \centering
    \includegraphics[width=\textwidth]{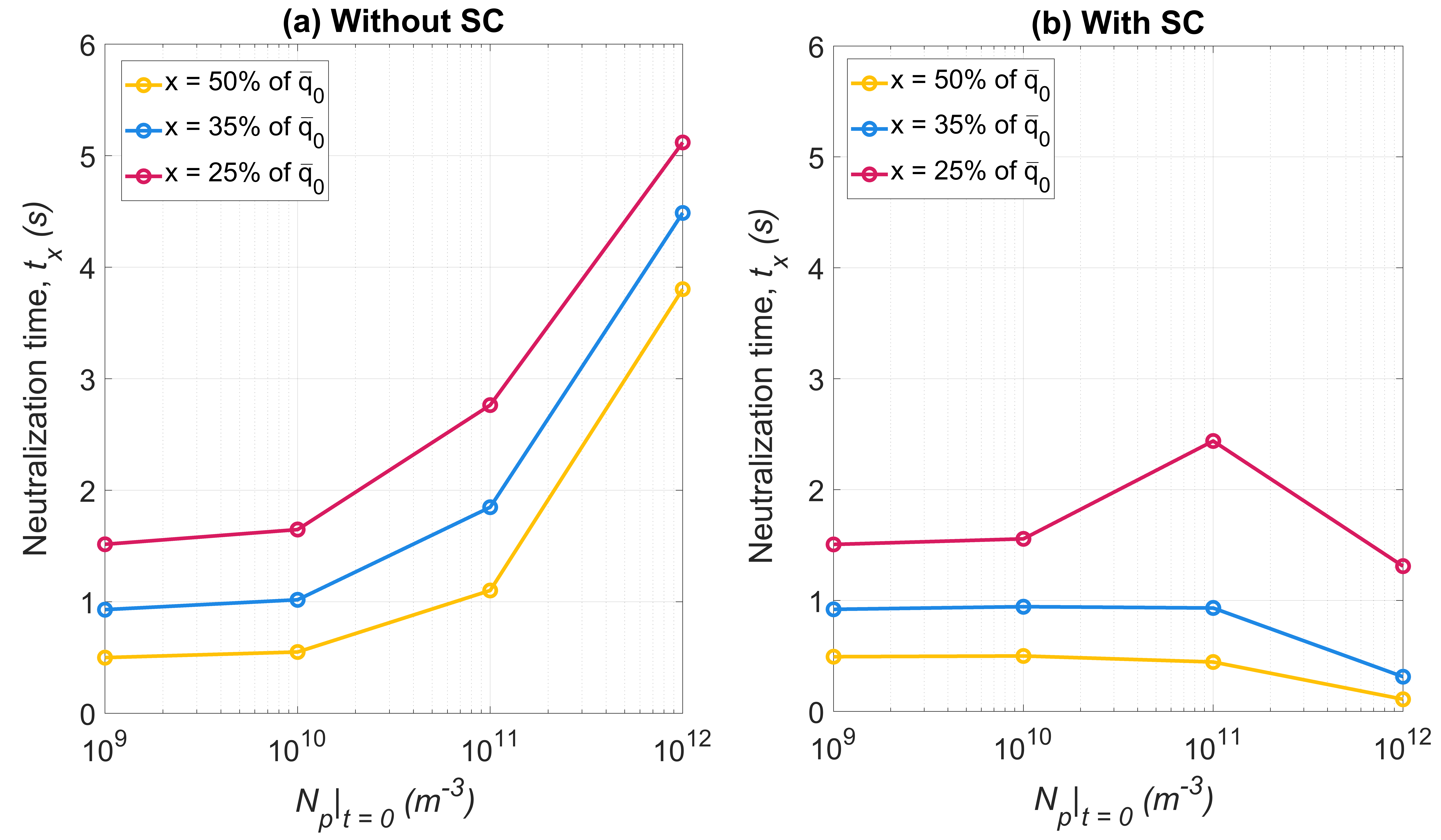}
    \caption{
       Neutralization time ($t_x$) required to reduce the initial mean charge per particle ($\bar q_0$) to $x$\% of its original value, where $x = 50$ (yellow), $35$ (blue), and $25$ (pink), as a function of the initial aerosol number concentration $N_p\big|_{t=0}$ (in m$^{-3}$), at a fixed ion concentration of $N_i = 10^{12}$~m$^{-3}$. Panel~(a) shows results without space charge (SC) effects, while panel~(b) includes SC. Simulations assume a monodisperse aerosol distribution with particle diameter $d_p = 1~\mu\mathrm{m}$.
}
    \label{fig:charge_reduction_vs_np}
\end{figure}

To quantitatively characterize the dynamics of aerosol charge neutralization, we introduce a new metric termed the \emph{neutralization time} ($t_x$), defined as the time required for the mean particle charge ($\bar{q}_0$), averaged over both the radial spatial domain and the full charge distribution, to decay to $x\%$ of its initial value. This measure captures not only the rate of charge dissipation but also incorporates the effects of ion–aerosol interactions, transport processes, and space charge dynamics. Figure~\ref{fig:charge_reduction_vs_np} shows the neutralization time $t_x$ for $x = 50$ (yellow), $35$ (blue), and $25$ (pink) as a function of the initial aerosol number concentration ($N_p\big|_{t=0}$), assuming a fixed ion concentration of $N_i = 10^{12}~\mathrm{m}^{-3}$, under both space charge (SC) and no-SC conditions.

The results clearly show that in the absence of space charge, the neutralization process becomes significantly slower with increasing aerosol concentration. This is expected, as a higher particle density leads to greater competition for available ions and reduced ion mobility, thereby impeding effective charge neutralization. However, when SC is included, the behavior changes markedly. The electric field induced by charge separation enhances ion drift toward oppositely charged particles, thereby accelerating the neutralization process. This enhancement is especially pronounced at higher aerosol concentrations, where space charge effects become dominant. A similar effect was reported by \citet{jidenko2022effect} in their study of mean particle charge under varying aerosol concentrations and SC conditions (see their Figure~6). They found that including SC led to significantly higher charging efficiency, particularly at elevated particle concentrations, due to electric field-assisted ion transport. This aligns with our findings, where SC substantially improves neutralization kinetics at high aerosol loads. Additionally, at high initial aerosol concentrations, it is plausible that space charge enhances radial electric field gradients, promoting faster particle migration toward the neutralizer walls. This could increase wall deposition rates (as shown in Fig.~\ref{aerosolconc}) and indirectly contribute to the observed sharp reduction in mean system charge.

\begin{figure}[H]
    \centering
    \includegraphics[width=\textwidth]{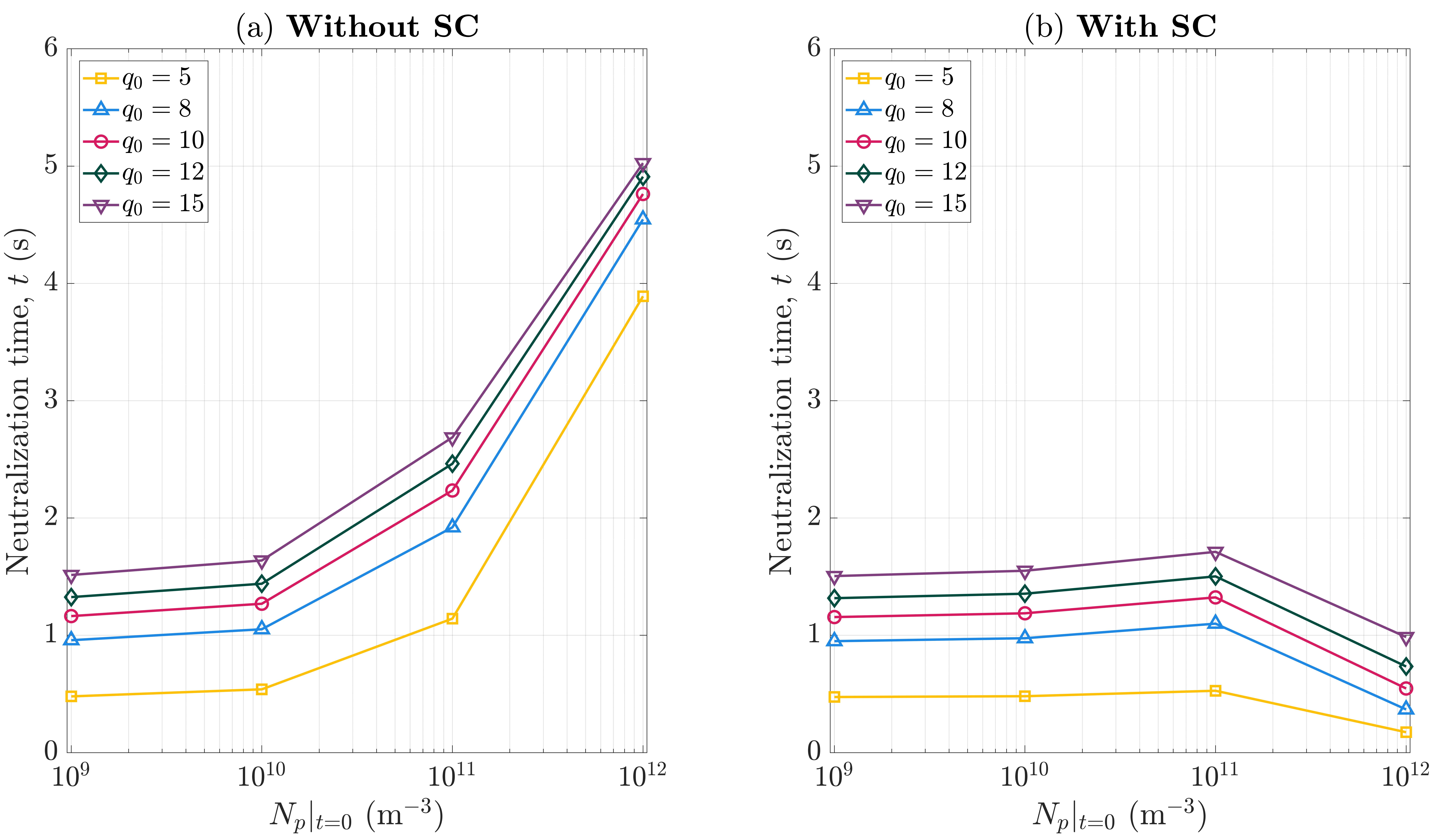}
    \caption{
       Neutralization time (\( t \)), defined as the time required to achieve mean charge per particle (\( \bar{q}=3 \))~e, as a function of the initial aerosol number concentration \( N_p\big|_{t=0} \) (in m\(^{-3}\)), for varying initial particle charge (\( q_0 \)). The ion concentration is fixed at \( N_i = 10^{12}~\mathrm{m}^{-3} \). Panel~(a) shows results without space charge (SC) effects, while panel~(b) includes SC. Simulations assume a monodisperse aerosol population with a particle diameter of \( d_p = 1~\mu\mathrm{m} \).
}
    \label{fig:charge_effect}
\end{figure}

Figure~\ref{fig:charge_effect} shows the neutralization time \( t \) required for the mean charge per particle to reach \( \bar{q} = 3~\mathrm{e} \), for different initial particle charges \( q_0 \), as a function of the initial aerosol number concentration \( N_p\big|_{t=0} \). Panel~(a) shows results without space charge (SC), while panel~(b) includes SC. 
Without SC, the neutralization time increases significantly with aerosol concentration and initial particle charge. Higher \( q_0 \) values result in longer times to reach \( \bar{q} = 3~\mathrm{e} \), especially at elevated \( N_p \), where competition for ions and ion depletion become more severe. This is expected, as particles with more initial charge require more ion attachments to reach the same target level of mean charge.

With SC included (panel~b), the behavior changes markedly. First, neutralization times are substantially reduced across all \( q_0 \) values, especially at higher \( N_p \). Second, the sensitivity of \( t \) to \( q_0 \) is much weaker, indicating that the SC-induced electric field compensates for the higher initial charge by enhancing ion drift and delivery to the particles. This explains the observed flattening and convergence of the curves at higher concentrations.
Interestingly, at \( N_p\big|_{t=0}  = 10^{12}~\mathrm{m}^{-3} \), the neutralization time decreases with increasing \( q_0 \) when SC is active, a reversal of the no-SC trend. This behavior may be linked to stronger space charge effects at higher initial system charge, which enhances the electric field and drives more efficient ion-particle interactions. The magnitude of \( q_0 \) therefore influences the strength of the induced field but not the qualitative trend, which remains consistent with the findings in Figure~\ref{fig:charge_reduction_vs_np}. These results reinforce the conclusion that SC dramatically accelerates neutralization under high aerosol loading, and that its benefits are increasingly pronounced with increasing particle charge.

\begin{figure}[H]
    \centering
    \includegraphics[width=\textwidth]{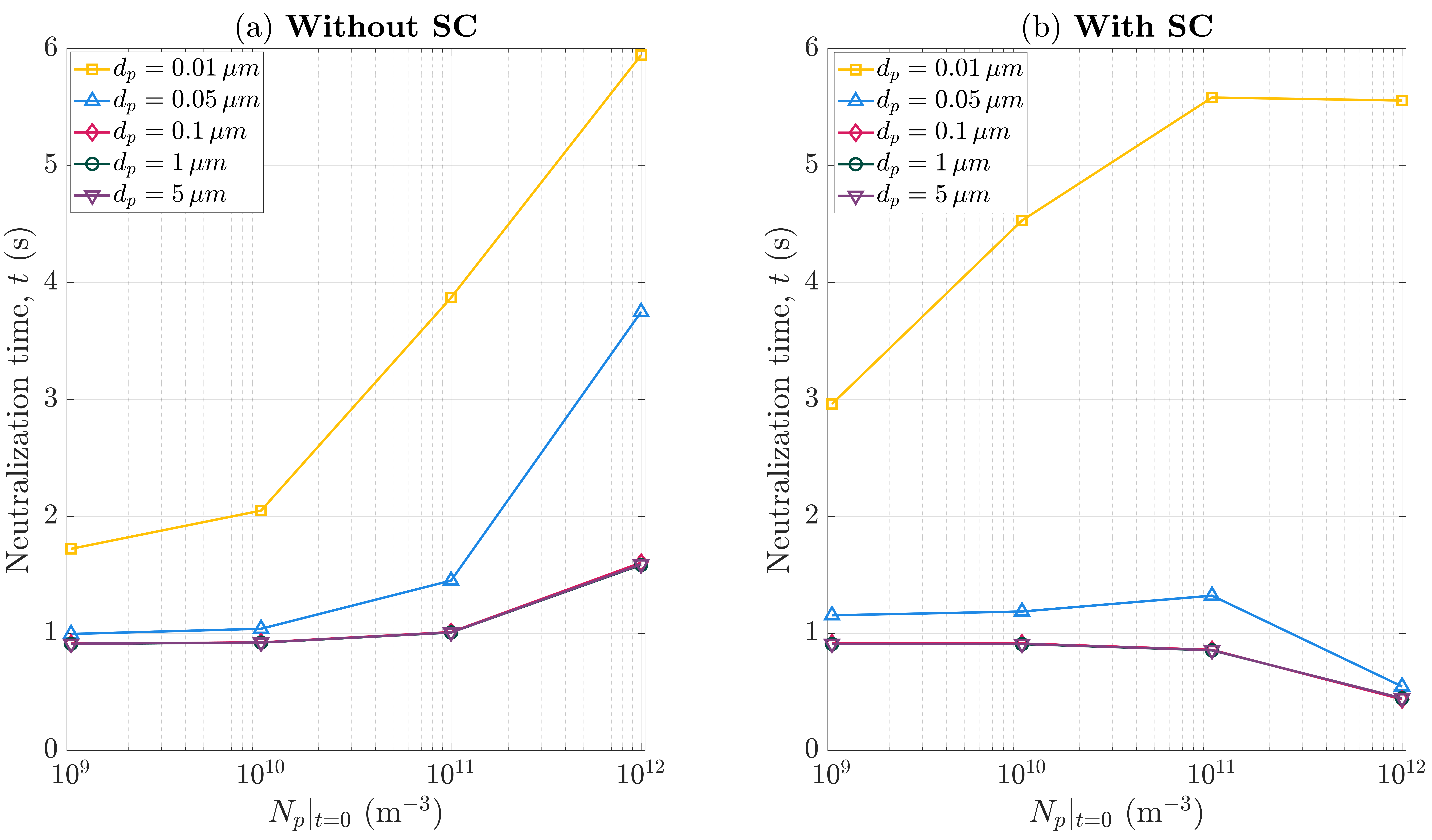}
    \caption{
       Neutralization time (\( t \)), defined as the time required to achieve mean charge per particle (\( \bar{q}=3 \))~e, shown as a function of the initial aerosol number concentration \( N_p\big|_{t=0} \) (in m\(^{-3}\)), for varying initial particle diameters \( d_p \). The ion concentration is fixed at \( N_i = 10^{12}~\mathrm{m}^{-3} \). Panel~(a) shows results without space charge (SC) effects, while panel~(b) includes SC. Simulations assume an initial particle charge of \( q_0 = +10~\mathrm{e} \).
}
    \label{fig:size_effect}
\end{figure}

Figure~\ref{fig:size_effect} shows the neutralization time \( t \) required for the mean particle charge to reach \( \bar{q} = 3~\mathrm{e} \), as a function of the initial aerosol number concentration \( N_p\big|_{t=0} \), for varying particle diameters \( d_p \). Panel~(a) presents results without space charge (SC) effects, while panel~(b) includes SC. All cases assume an initial particle charge of \( q_0 = +10~\mathrm{e} \) and a fixed ion concentration of \( N_i = 10^{12}~\mathrm{m}^{-3} \).
In the absence of SC, a strong size dependence is observed. Smaller particles (\( d_p = 0.01~\mu\mathrm{m} \)) exhibit significantly longer neutralization times, especially at higher aerosol concentrations. This is attributed to their smaller surface area and weaker electrostatic interactions, which reduce the likelihood and rate of ion attachment under identical ion and flow conditions. As particle size increases, the surface area available for charge accumulation also increases, and stronger electric fields at the particle surface promote more efficient ion capture, resulting in shorter neutralization times.

When SC is included (panel~b), neutralization times are reduced for all sizes, and the size dependence is substantially suppressed. The SC-induced electric field enhances ion drift toward the particles, offsetting the natural disadvantage that smaller particles have due to their low surface area. As a result, neutralization becomes more efficient and less sensitive to particle size. For \( d_p \geq 0.1~\mu\mathrm{m} \), the effect of size on neutralization time is almost negligible when SC is present, and the curves for larger sizes collapse toward a common value. However, the smallest particles still neutralize more slowly than others even under SC conditions, likely because their inherently low ion capture efficiency cannot be fully compensated by enhanced ion transport.

At high aerosol concentrations (\( N_p\big|_{t=0} = 10^{12}~\mathrm{m}^{-3} \)), the SC effect is particularly pronounced. For example, the difference in neutralization time between \( d_p = 0.01~\mu\mathrm{m} \) and \( d_p = 1~\mu\mathrm{m} \) is dramatically reduced compared to the no-SC case, demonstrating the robustness of SC-assisted ion delivery under crowded conditions.

Overall, these results highlight that SC not only accelerates neutralization but also acts as an equalizing mechanism across particle sizes, mitigating the limitations faced by smaller aerosols in capturing ions effectively.

\section{Discussion}
The classical $N_i t$ product has long served as a practical metric for estimating aerosol charge neutralization in dilute systems. It assumes a single characteristic timescale for the decay of particle charge and relies on spatially uniform ion concentrations and low aerosol loading. However, the results presented here demonstrate that this approach breaks down under realistic conditions where fixed ion generation becomes insufficient to match increasing aerosol charging demand. In such cases, ion depletion, space charge effects, and nonlinear coupling between ions and aerosols significantly alter the relaxation dynamics.

Our simulations show that, particularly in high concentration aerosol environments, neutralization cannot be described by a single exponential decay governed by a fixed $N_i t$ threshold. Instead, both the ion concentration and the mean particle charge evolve together, introducing multiple interacting timescales. To address this, we derive an analytical expression for the mean charge decay that accounts for the coupled evolution of ion and aerosol populations.

The resulting solution, derived in detail in the Supplementary Information (Section~\ref{SI:relaxation}), provides a generalized framework for estimating charge relaxation under finite aerosol concentrations:
\begin{equation}
\bar{q}(t) = q_0 e^{-\lambda_2 t} + \frac{\beta_0 N_p q_0}{\lambda_2 - \lambda_0} \left( e^{-\lambda_0 t} - e^{-\lambda_2 t} \right),
\label{eq:qmain}
\end{equation}
where \( q_0 \) is the initial mean charge, \( N_p \) is the aerosol number concentration, \( \beta_0 \) is the ion–aerosol attachment coefficient, and the time constants are defined as
\[
\lambda_0 = 2 \nu_0 N_{i,\infty}, \qquad \lambda_2 = \beta_0 N_p + \nu_0 N_{i,\infty}.
\]

This expression captures a two-timescale decay: an initial rapid neutralization driven by direct attachment, followed by a slower relaxation modulated by evolving space charge. In the dilute limit \( N_p \to 0 \), Eq.~\eqref{eq:qmain} reduces to a single exponential form consistent with classical theory. However, as aerosol concentration increases, both terms play a significant role, and the traditional $N_i t$ product fails to describe the neutralization dynamics accurately. This expression also provides the theoretical basis for the neutralization time metric $t_x$ introduced earlier, which corresponds to the time at which $\bar{q}(t)$ decays to a specified fraction of its initial value. In this way, the $t_x$ curves presented in Figure~\ref{fig:charge_reduction_vs_np} provide a direct visualization of the charge relaxation behavior encoded in Eq.~\eqref{eq:qmain}.

We therefore propose that Eq.~\eqref{eq:qmain} serves as a more general alternative to the $N_i t$ product, particularly in the design and analysis of systems involving dense aerosol flows or weak ion sources. This model offers improved predictive capability and clarifies the role of space charge in enabling complete neutralization in otherwise symmetric ion environments.

\section{Conclusion}

This study presents a detailed analysis of aerosol charge neutralization under conditions where both ion generation and aerosol concentrations are finite, which are conditions commonly encountered in laboratory, atmospheric, and industrial systems. While classical neutralizer theory relies on the \( N_i t \) product as a universal metric for charge decay, our results show that this assumption breaks down when space charge (SC) effects are included.

Using a coupled ion–aerosol transport model that self-consistently solves for internal electric fields, we demonstrate that even small net charge densities can generate SC fields strong enough to significantly alter ion fluxes. This results in enhanced neutralization near the centerline and ion depletion near the boundary. This phenomena is entirely missed in conventional models that assume uniform ion distributions. These deviations grow stronger with increasing aerosol concentration and cannot be captured by the classical single-timescale exponential decay model.

To account for these effects, we derived a generalized analytical expression for the time evolution of mean particle charge, \( \bar{q}(t) \), which includes the coupled influences of ion depletion, aerosol loading, and SC-induced drift. The resulting two-timescale decay model recovers the classical \( N_i t \) behavior in the dilute limit but predicts significant deviations in dense systems. This generalized form provides a mechanistic explanation for why traditional neutralization thresholds (e.g., $50\%$ decay) become increasingly unreliable at higher aerosol concentrations.

A key outcome of this work is the introduction of a diagnostic metric, neutralization time \( t \), used to systematically evaluate how aerosol number concentration \( N_p \), initial particle charge \( q_0 \), and particle size \( d_p \) influence neutralization dynamics. In the absence of SC, \( t \) increases significantly with both \( N_p \) and \( q_0 \), as higher particle loading and charge levels intensify ion competition, thereby slowing charge decay. Smaller particles also exhibit longer neutralization times due to their lower surface area and weaker electrostatic attraction for ions.
At high aerosol concentrations, SC effects become the primary driver of charge relaxation, sharply reducing neutralization times and promoting more uniform behavior across a range of particle sizes and initial charges.

These insights clarify the limitations of classical neutralizer theory and offer a physically grounded framework for evaluating charge relaxation in real systems. The analytical model and diagnostic tools developed here, particularly the neutralization time metric, are directly relevant to the design, interpretation, and optimization of neutralization devices operating under non-ideal, space charge–affected conditions. More broadly, this work highlights the essential role of space charge in governing aerosol–ion dynamics, with implications for a wide range of electrostatic processes involving charged particles and confined flows.

\section{Acknowledgment} 
 This work was supported by institutional resources at the University of Leeds and Indian Institute of Technology Bombay. No external funding was received. 

 \section{Author Contributions}
YSM conceived the idea and built the initial framework. RP, YSM and KG set up the analytical formulations. KG developed the numerical version and implemented the model. KG did the calculations and analysed the results with contributions from YSM and GS. KG, GS, RP and YSM wrote the manuscript. All authors approved the final text.

\section{Supplementary information}

\subsection{Electric field}
\begin{figure}[H]
    \centering
    \includegraphics[width=\textwidth]{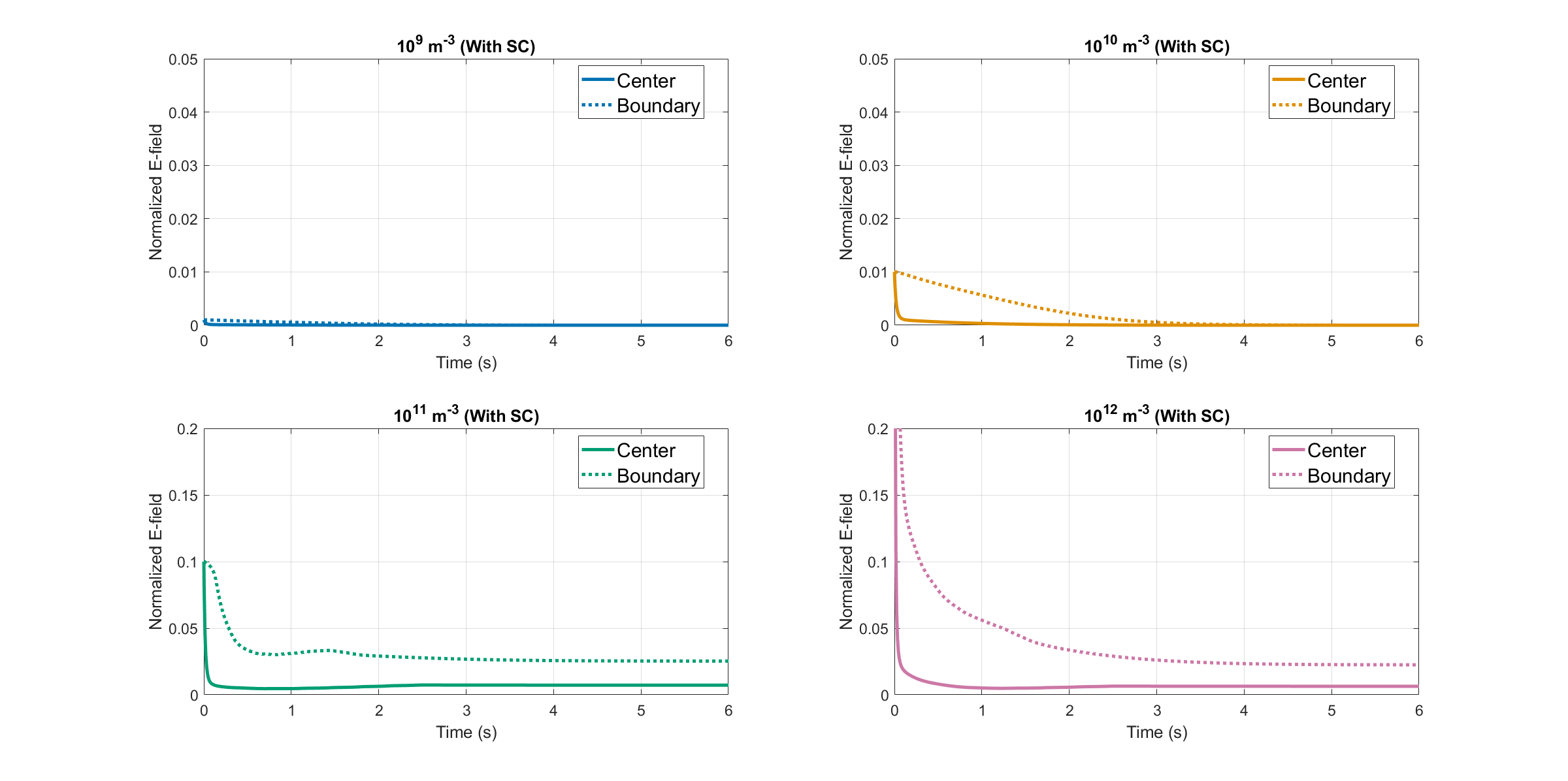}
    \caption{Time evolution of normalized electric field at the center and boundary of the neutralizer for different aerosol number concentrations ($10^{9}$ to $10^{12}$~m$^{-3}$) under space charge (SC) conditions. Each subplot corresponds to a specific concentration level,showing the distinct electric field response at both spatial locations. Electric field strength is higher and more persistent near the boundary, particularly at higher concentrations due to accumulated space charge effects.}
    \label{fig:normalized_efield}
\end{figure}

\subsection{Velocity profile}
\begin{figure}[H]
    \centering
    \includegraphics[width=\textwidth]{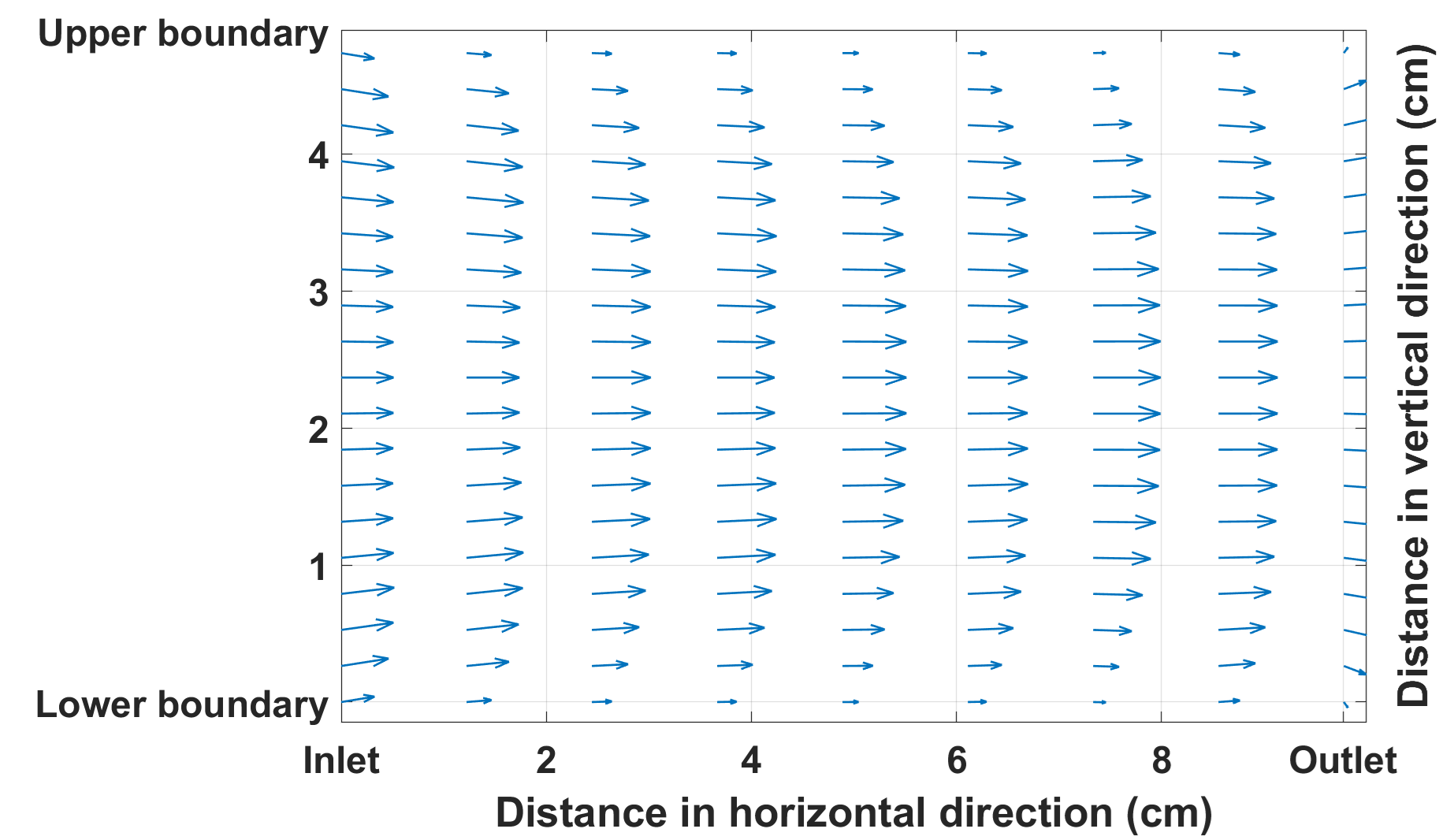}
    \caption{Simulated velocity profile for laminar air flow at 2 L/min through a cylindrical neutralizer, obtained by solving the steady-state incompressible Navier–Stokes equations in a two-dimensional axisymmetric domain. The arrows represent the local velocity vectors. The parabolic profile reflects fully developed laminar flow, with maximum velocity along the central axis and no-slip boundary conditions enforced at the upper and lower walls.}
    \label{fig:velocity profile}
\end{figure}

\subsection{Key role of  space charge in neutralization process}
\label{SI}
In their now classic papers, Hoppel and Frick~\citep{hoppel1986ion,hoppel1990nonequilibrium} examined the issue of symmetric and asymmetric Boltzmann distributions from the perspective of the charge neutrality condition combined with the bipolar ion-aerosol attachment process. Considering a spatially homogeneous system, they argued that if the ion properties are perfectly symmetric, then an originally charged aerosol system will attain zero mean charge, leading to a symmetric Boltzmann distribution. The key to this result is the assumption that the combined system of ions and aerosols would attain perfect charge neutrality in space after a sufficiently long time.

However, even if the ions are completely symmetric, the mean charge of an initially charged aerosol system does not decay to zero through attachment processes alone. One must necessarily include the space-charge effect, however small, to ensure complete neutralization. This result highlights the critical role of space charge in the neutralization process: although often considered negligible, it becomes essential for achieving complete charge relaxation under realistic aerosol loading.

To illustrate this, suppose an aerosol is spatially distributed uniformly at a concentration $N_{p}$ (particles per unit volume), maintained in a bipolar ion environment generated by a radioactive neutralizer with ion production rate density $S$. We show that even if the ions are perfectly symmetric in respect of their mobility characteristics, an initially charged aerosol will not be neutralized unless we incorporate space charge effects.  In other words, the space-charge-induced drift is the guarantor of complete neutralization of charged aerosol in a symmetric bipolar ion system and not merely the symmetry of the ions. 

Because of the assumption of spatially homogeneity, the gradient effects of convection and diffusion may be ignored. The ion balance equations, excluding space-charge effect, but including ion asymmetry, production, recombination and attachment to aerosols, are given below:

\begin{align}
\frac{dN_{i}^+}{dt} &= S - \alpha_0 N_{i}^+ N_{i}^- - N_{i}^+ \sum_{q=-\infty}^{\infty} \beta_q^+  N_{p}^q, \label{eq:Ni_plus} \\
\frac{dN_{i}^-}{dt} &= S - \alpha_0 N_{i}^+ N_{i}^- - N_{i}^- \sum_{q=-\infty}^{\infty} \beta_q^-  N_{p}^q, \label{eq:Ni_minus} \\
\frac{d N_{p}^q}{dt} &= N_{i}^+ \beta_{q-1}^+  N_{p}^{q-1} + N_{i}^- \beta_{q+1}^-  N_{p}^{q+1} - (N_{i}^+ \beta_q^+ + N_{i}^- \beta_q^-)  N_{p}^q. \label{eq:Np_q}
\end{align}

The space charge density $\rho$ is defined as:
\begin{equation}
\rho = N_{i}^+ - N_{i}^- + \bar{q} N_{p} \label{eq:rho_def}
\end{equation}

The mean charge ($\bar{q}$) per particle is:
\begin{equation}
\bar{q} = \frac{1}{N_{p}} \sum_{q=-\infty}^{\infty} q  N_{p}^q \label{eq:qbar_def}
\end{equation}
where $N_{p}=\sum_{q=-\infty}^{\infty} N_{p}^q$ is the total (all charges) particle concentration. 
From Eqs.~\eqref{eq:Np_q} and~\eqref{eq:qbar_def},  the evolution of mean charge can be shown to satisfy the following equation:
\begin{equation}
\frac{d(\bar{q} N_{p})}{dt} = \sum_{q=-\infty}^{\infty} (N_{i}^+ \beta_q^+ - N_{i}^- \beta_q^-)  N_{p}^q \label{meanchrgeaero}
\end{equation}
For illustration purposes, we restrict ourselves to the case of continuum attachment coefficients:
\[
\beta_q^+ = \frac{\beta_0^+ q \alpha e^{-q\alpha}}{1 - e^{-q\alpha}}, \quad
\beta_q^- = \frac{\beta_0^- q \alpha}{1 - e^{-q\alpha}},
\]
where $\alpha = r_c / a$ and $r_c = e^2 / (4 \pi \varepsilon_0 k_B T)$ is the Coulomb length, and $\beta_0^\pm = 4 \pi a D^\pm$ are the attachment coefficients between the neutral particles and positive \& negative ions.

An equation for the space charge density $\rho$, defined by Eq.~\eqref{eq:rho_def} is obtained by carefully working through Eqs.~\eqref{eq:Ni_plus} to \eqref{eq:Np_q} and Eqs.~\eqref{eq:rho_def} to \eqref{meanchrgeaero}. While doing so, we note that the charging terms cancel exactly and we obtain:
\begin{equation}
\frac{d\rho}{dt} = 0
\label{rhochange}
\end{equation}
The solution to eq \eqref{rhochange} is $\rho(t)$=constant at all times.  If originally the particle carried a charge $q_0$, and the original ion concentrations were equal $N_{i}^+ (0)= N_{i}^- (0)$, then Eq.\eqref{rhochange}  implies that 
\begin{equation}
\rho(t) = \text{constant} = \rho_0 = N_{p} q_0 \ \label{eq:rho_const}
\end{equation}
As this equation is true at all times, it should be true in steady-state ($t \to \infty$) as well. Hence in steady-state, 
\begin{equation}
N_{i,\infty}^+ - N_{i,\infty}^- + N_{p} \bar{q}_\infty = N_{p} q_0 \label{eq:steady_charge}
\end{equation}

Within the continuum model, the steady-state solution of Eq.~\eqref{eq:Np_q} yields a shifted Boltzmann charge distribution with:
\begin{equation}
\bar{q}_\infty = \frac{1}{\alpha} \left[ \ln\left( \frac{N_{i,\infty}^+}{N_{i,\infty}^-} \right) + \ln\left( \frac{\mu^+}{\mu^-} \right) \right] \label{eq:qbar_boltzmann}
\end{equation}

Then the solution for the steady-state mean charge of aerosol will be a solution of the following transcendental equation 
\begin{equation}
\bar{q}_\infty = \frac{1}{\alpha} \ln\left( \frac{\mu^+}{\mu^-} \cdot \frac{2N_{i,\infty} + N_{p}(q_0 - \bar{q}_\infty)}{2N_{i,\infty} - N_{p}(q_0 - \bar{q}_\infty)} \right) \label{eq:transcendental}
\end{equation}
Where $N_{i,\infty}$ is the sign independent average ion concentration 
\begin{equation}
N_{i,\infty} = \frac{1}{2} (N_{i,\infty}^+ + N_{i,\infty}^-) \label{eq:Ninf}
\end{equation}

Eq.~\eqref{eq:transcendental} requires a numerical solution for $\bar{q}_\infty$. To illustrate that Eq.~\eqref{eq:Ninf} yields a nonzero solution for $\bar{q}_\infty$ even when $\mu^+ = \mu^-$, we assume the high ion density limit, i.e., $N_{i,\infty} \gg N_{p}$. Then, the terms within the logarithmic function can be linearized, and one obtains,

\begin{equation}
\bar{q}_\infty = \frac{1}{\alpha} \left[ \frac{N_{p}}{N_{i,\infty}} (q_0 - \bar{q}_\infty) + \ln \gamma \right], \quad \gamma = \frac{\mu^+}{\mu^-} \label{eq:linearized}
\end{equation}

Eq.\eqref{eq:linearized} possesses a solution 
\begin{equation}
\bar{q}_\infty = \frac{\frac{N_{p}}{N_{i,\infty}} q_0 + \ln \gamma}{\alpha + \frac{N_{p}}{N_{i,\infty}}} \label{eq:qbar_solution}
\end{equation}

From this we see that even if $\gamma = 1$, (perfectly symmetric ions), $\bar{q}_\infty$ does not decay to zero, but remains at a steady fraction of the original charge:

\begin{equation}
\bar{q}_\infty(\gamma = 1) = \frac{q_0}{1 + \frac{\alpha N_{i,\infty}}{N_{p}}} \label{eq:qbar_gamma1}
\end{equation}

This counter-intuitive result, namely persistence of mean charge on a spatially homogeneous, initially charged aerosol, even in perfectly symmetric ion system, points to a fundamental inconsistency in the classical understanding of neutralization. This apparent anomaly can only be resolved by explicitly incorporating space-charge–induced drift effects, as we demonstrate below.  
The phenomenon of space charge induced drift does not require spatial gradients in concentrations or charge densities, and hence is consistent with the assumption of spatially homogeneous system. When we introduce space-charge induced drift, Eqs.\eqref{eq:Ni_plus}, \eqref{eq:Ni_minus} get modified as follows. 

\begin{align}
\frac{dN_{i}^+}{dt} &= -\mu^+ N_{i}^+ \nabla \cdot \mathbf{E} + S - \alpha N_{i}^+ N_{i}^- - N_{p} N_{i}^+ \sum_q \beta_q^+ f_q, \label{qpluswithSC} \\
\frac{dN_{i}^-}{dt} &= \mu^- N_{i}^- \nabla \cdot \mathbf{E} + S - \alpha N_{i}^+ N_{i}^- - N_{p} N_{i}^- \sum_q \beta_q^- f_q, \label{qminuswithSC}
\end{align}

and the charging equation becomes:
\begin{equation}
\frac{d N_{p}^q}{dt} = -q e B  N_{p}^q \nabla \cdot \mathbf{E} +  N_{i}^+ \beta_{q-1}^+  N_{p}^{q-1} + N_{i}^- \beta_{q+1}^-  N_{p}^{q+1} - (N_{i}^+ \beta_q^+ + N_{i}^- \beta_q^-)  N_{p}^q. \label{aerosolcharge}
\end{equation}

Where $B$ is the mechanical mobility of the particle. The Poisson’s equation in air for the space-charge induced electric field is 

\begin{equation}
\nabla \cdot \mathbf{E} = \frac{e}{\varepsilon} \rho 
\label{spacecharge}
\end{equation}

The time rate of change of the space charge density $\rho$  defined by Eq. \eqref{eq:rho_def} is no longer zero as in Eq. \eqref{rhochange}; instead, it assumes the form

\begin{equation}
\frac{d\rho}{dt} = -\left( \mu^+ N_{i}^+ + \mu^- N_{i}^- + N_{p} \bar{q}^2 e B \right) \nabla \cdot \mathbf{E} 
\label{chargedensitynew}
\end{equation}

Upon replacing $\mathbf{E}$ in Eq.\eqref{chargedensitynew} using Eq.\eqref{spacecharge}, we obtain

\begin{equation}
\frac{d\rho}{dt} = -\left( \mu^+ N_{i}^+ + \mu^- N_{i}^- + N_{p} \bar{q}^2 e B \right) \frac{e}{\varepsilon} \rho 
\label{chargedensitynew1}
\end{equation}

Since the coefficient of the rhs will always be positive non-zero regardless of whether ions are symmetric or otherwise because so long as ion productions exists, ion concentrations will persist. Then Eq. \eqref{chargedensitynew1} implies that the space charge will decay exponentially to zero. i.e.

\begin{equation}
\rho(t) = \rho_0 \exp\left[ -\left( \mu^+ N_{i}^+ + \mu^- N_{i}^- + N_{p} \bar{q}^2 e B \right) \frac{e}{\varepsilon} t \right] 
\label{chargedensitynew2}
\end{equation}

Hence the final space-charge density $\rho(t \to \infty)=\rho_\infty =0$ regardless of the initial space-charge.  This is equivalent to the assumption of quasi-neutrality condition, a key result that is guaranteed only with the inclusion of space-charge effect. With this, from Eq. \eqref{eq:rho_def}, we have 

\begin{equation}
\bar{q}_\infty = -\frac{1}{N_{p}}(N_{i,\infty}^+ - N_{i,\infty}^-) \label{eq:qbar_final}
\end{equation}

As in Eq.\eqref{eq:qbar_boltzmann}, second relationship for the mean charge $\bar{q}_\infty$, follows from the steady-state charge distribution, including ion asymmetry ($\gamma$):  
\begin{equation}
\bar{q}_\infty = \frac{1}{\alpha} \left[ \ln\left(\frac{N_{i,\infty}^+}{N_{i,\infty}^-}\right) + \ln \gamma \right], \quad \text{where } \alpha = \frac{r_c}{a} 
\label{qbarinfy}
\end{equation}

To solve Eqs.\eqref{eq:qbar_final}, \eqref{qbarinfy} , we introduce quantities representing net ion density ($\phi$) and total ion density ($2N_{i}$) as follows:  
\begin{align}
\phi &= \frac{N_{i,\infty}^+ - N_{i,\infty}^-}{N_{i,\infty}^+ + N_{i,\infty}^-} \label{iondensity} \\
N_{i,\infty} &= \frac{N_{i,\infty}^+ + N_{i,\infty}^-}{2} \label{totaliondensity}
\end{align}

With Eqs.(\eqref{iondensity},\eqref{totaliondensity}), the Eqs.(\eqref{eq:qbar_final},\eqref{qbarinfy}) transform to  
\begin{equation}
\bar{q}_\infty = -\frac{2N_{i,\infty}}{N_p} \phi 
\label{qbarinfy1}
\end{equation}
and  
\begin{equation}
\bar{q}_\infty = \frac{1}{\alpha} \left[ \ln\left( \frac{1+\phi}{1-\phi} \right) + \ln \gamma \right] \approx \frac{1}{\alpha} \left[ 2\phi + \ln \gamma \right] \quad \text{for weak ion asymmetry } (\phi \ll 1) 
\label{qbarinfy2}
\end{equation}

Upon eliminating $\phi$, the solution to Eqs.(\eqref{qbarinfy1},\eqref{qbarinfy2}) is  
\begin{equation}
\bar{q}_\infty = \frac{\ln \gamma}{\alpha + \frac{N_{p}}{N_{i,\infty}}} 
\label{qbarinfy_final}
\end{equation}

This expression is identical to the $\gamma$-dependent part of the mean charge expression obtained without invoking space-charge effect (Eq. \eqref{eq:qbar_solution}). It however, gets rid of the persistence of original charge ($q_0$). As a result, for perfectly symmetric system of ions \{$\gamma = \mu^+ / \mu^- = 1$\}, unlike Eq.~(\eqref{eq:qbar_solution},\eqref{eq:qbar_gamma1}), it follows from Eq.\eqref{qbarinfy_final} that $\bar{q}_\infty = 0$, i.e., the aerosol system will eventually undergo complete neutralization regardless of the relative concentrations of particles and ions. The introduction of space-charge effect at a conceptual level guarantees neutralization for an ideal situation of symmetric ions. In a realistic case of ion asymmetry ($\gamma \ne 1$), a nonzero value of aerosol mean charge $\bar{q}_\infty$ persists, which decreases gradually as $N_{p}/N_{i,\infty} \to \infty$. These matters are consistent with the results of  \cite{hoppel1990nonequilibrium}.

\subsection{Aerosol concentration dependent relaxation of particle mean charge}
\label{SI:relaxation}
The classical $N_{i}t$ product metric assumes dilute aerosol conditions and a single exponential decay in mean particle charge. However, under realistic conditions involving high particle concentrations and limited ion production, neutralization dynamics become significantly more complex. This section presents an analytical framework for describing the relaxation of particle mean charge as a function of aerosol concentration, capturing the transition from single- to dual-timescale behavior.

In the dilute aerosol limit and under continuum conditions, the decay of the mean particle charge and the consequent relaxation of the aerosol population to Boltzmann equilibrium occurs exponentially as:
$e^{-\nu_0 N_{i} t}$, where, 
\begin{equation}
\nu_0 = 4 \pi r_c D_i
\end{equation}
is the neutralization coefficient (in $\mathrm{m^3\,s^{-1}}$) in the continuum regime, $D_i$ is the diffusion coefficient of ions, and $N_{i}$ is the ion concentration, defined as the average number density of either positive or negative ions. The coefficient $\nu_0$ increases by approximately a factor of two as the system approaches the free molecular regime.
The near constancy of $\nu_0$ in the continuum regime, with a typical value of $2.46 \times 10^{-12}~\mathrm{m^3/s}$, leads to the concept of a typically required $N_{i}t$ product of approximately $10^{13}~\mathrm{s/m^3}$ for complete neutralization.
However in general there is both size dependency as well as concentration dependency for the $\nu_0$.  This aspect has been investigated for unipolar chargers in great detail \citep{jidenko2022effect}. In the context of neutralization, the size dependency has been examined extensively in the past: Here we present analytical approximation for the particle concentration dependence of the neutralization coefficient. These approximations are based on asymptotic linearization of the charging equations  assuming small excess mean charge on particles. The formulae will be helpful for order estimates of the time required for neutralization in realistic case of non-zero aerosol concentration and weak ion sources. 
As we saw in Section \ref{SI}, because of the slight asymmetry in ionic mobilities, the original aerosol charge does not decay to zero but asymptotically approaches a finite charge given by 
\begin{equation}
\bar{q}_\infty = \frac{\ln \gamma}{\alpha + \dfrac{N_{p}}{N_{i,\infty}}}
\end{equation}

This is shown in Eq. \ref{qbarinfy_final} of Section \ref{SI}. However, for relaxation rate estimation, the ion asymmetry is of secondary consequence, the primary effect being that of the mean mobilities,  ion concentrations  and aerosol attachments.  In view of this, for rate estimation, we may invoke symmetric situations, i.e. $\mu^+ = \mu^- = \mu$
Also we consider large volume case so that spatial concentration gradients are insignificant and the surface removal effects are ignored.

\subsubsection{Ion concentration profiles}
Let us define the following notations:

\begin{align}
n_d(t) &= \frac{1}{2} \left[ N_{i}^+(t) - N_{i}^-(t) \right] \quad \text{(ion concentration difference)} \label{eq:nd} \\
N_{i}(t)   &= \frac{1}{2} \left[ N_{i}^+(t) + N_{i}^-(t) \right] \quad \text{(sign-averaged one-component ion concentration)} \label{eq:N}
\end{align}

\noindent
Omitting the time dependence for brevity, we have:
\begin{align}
N_{i}^+ &= N_{i} + n_d \\
N_{i}^- &= N_{i} - n_d
\end{align}

\noindent
In practice, the ion concentration difference is much smaller than the average ion concentration, i.e., $n_d \ll N_{i}$.

We ignore all terms of order higher than linear in $n_d$ or $\bar{q}$. Hence ion balance equations for sign averaged one component concentration (see Section \ref{SI}) is:
\begin{equation}
\frac{dN_{i}}{dt} = S - \alpha_0 N_{i}^2 - \beta_0 N_{p} N_{i}
\end{equation}

Let the time start at the point of injecting aerosol into the system until which ions were in a steady-state concentration controlled purely by recombination: i.e.
\begin{equation}
N_{i}(0) = N_{i,0} = \sqrt{\frac{S}{\alpha_0}}
\end{equation}
Due to attachment to aerosols, a new steady-state will be reached. In this steady-state the final average ion concentration, with aerosol, is
\begin{equation}
N_{i,\infty} = N_{i,0} \left( \sqrt{k^2 + 1} - k \right) \approx N_{i,0} (1 - k), \quad \text{for } k \ll 1
\end{equation}
where, 
\begin{equation}
k = \frac{\beta_0 N_{p}}{2 \sqrt{\alpha_0 S}}.
\end{equation}
$k$ is the factor that represents aerosol concentration effect on averaged one-component ion density.  The final steady-state ion concentration decreases with increasing $N_{p}$ (or $k$) due to dominance of ion- loss by attachment to aerosol particles.

The time relaxation of $N_{i}$ from its initial value $N_{i,0}$ to the steady-state value $N_{i,\infty}$ is nonlinear. To estimate it to a linear order, let
\begin{equation}
N_{i}(t) = N_{i,\infty} + N'_{i}(t).
\end{equation}
Then,
\begin{align}
\frac{dN_{i}'}{dt} &= S - \left[ \alpha_0 \left(N_{i,\infty}^2 + 2N_{i,\infty} N_{i}'(t) + {N_{i}'}^2 \right) + \beta_0 N_{p} \left(N_{i,\infty} + N_{i}'(t)\right) \right] \notag \\
&= - (2 \alpha_0 N_{i,\infty} + \beta_0 N_{p}) N_{i}'(t) - \alpha_0 {N_{i}'}^2
\end{align}
We neglect the quadratic term $\alpha_0 {N_{i}'}^2$ under the assumption that $N_{i}'(t) \ll N_{i,\infty}$, yielding the linearized form:
\begin{align}
\frac{dN_{i}'}{dt} &= -\left( 2 \alpha_0 N_{i,\infty} + \beta_0 N_{p} \right) N'_{i}(t) \notag \\
&= -\left[ 2 \sqrt{\alpha_0 S} \left( \sqrt{k^2 + 1} - k \right) + 2k \sqrt{\alpha_0 S} \right] N'_{i}(t) \notag \\
&= -2 \sqrt{\alpha_0 S} \sqrt{k^2 + 1} \, N'_{i}(t),
\end{align}

shows that it approaches an exponential relaxation asymptotically in the form:
\begin{equation}
N_{i}(t) = N_{i,\infty} + (N_{i,0} - N_{i,\infty}) e^{-\lambda_1 t}
\end{equation}
where the relaxation rate for total concentration is
\begin{equation}
\lambda_1 = 2 \sqrt{\alpha_0 S + \frac{1}{4} \beta_0^2 N_{p}^2} = 2 \sqrt{\alpha_0 S} \, \sqrt{1 + k^2},
\end{equation}
and the corresponding relaxation time is given by
\begin{equation}
t_{c1} = \frac{1}{\lambda_1}.
\end{equation}
Therefore, aerosol loading decreases the time required for the ion concentration to relax to steady state.

\subsubsection{Relaxation of particle mean charge}

We refer to Eq. \ref{aerosolcharge} in Section \ref{SI}.
\begin{equation}
N_{p} \frac{d\bar{q}}{dt} = \sum_{q=-\infty}^{\infty} \left( N_{i}^+ \beta_q^+ - N_{i}^- \beta_q^- \right) N_{p}^q = \sum_{q=-\infty}^{\infty} \left[ (N_{i} + n_d)\beta_q^+ - (N_{i} - n_d)\beta_q^- \right]  N_{p}^q
\end{equation}
We linearize as: \[
\beta_q^{\pm} \approx \beta_0 \left( 1 \mp \frac{\alpha q}{2} \right).
\]
Ignoring higher order terms, linearization of the charging eq leads to:
\begin{equation}
\frac{d\bar{q}}{dt} = -\beta_0 \alpha N_{i} \bar{q} + 2 \beta_0 n_d, \quad \text{where } \alpha = \frac{r_c}{a}
\end{equation}
Eliminating $n_d$ using the definition of spatial charge density, namely
\[
\rho = N_{i}^+ - N_{i}^- + \bar{q} N_{p} = 2 n_d + \bar{q} N_{p},
\]
we obtain:
\begin{equation}
\frac{d\bar{q}}{dt} = -(\beta_0 N_{p} + \nu_0 N_{i}) \bar{q} + \beta_0 \rho = -\left[ \beta_0 N_{p} + \nu_0 N_{i,\infty} + \nu_0 (N_{i,0} - N_{i,\infty}) e^{-\lambda_1 t} \right] \bar{q} + \beta_0 \rho
\end{equation}
In Section \ref{SI}, we have shown that the space charge density $\rho = 2n_d + N_p\bar{q}$ satisfies the following exact equation:
\begin{equation}
\frac{d\rho}{dt} = -\frac{2 N_{i} \mu e}{\varepsilon} \rho = -2 \nu_0 \left[ N_{i,\infty} + (N_{i,0} - N_{i,\infty}) e^{-\lambda_1 t} \right] \rho
\label{eq:rho_evolution}
\end{equation}
where
\[
\nu_0 = \frac{\mu e}{\varepsilon} = \frac{e^2 D_i}{\varepsilon k_B T} = 4 \pi r_c D_i = \alpha \beta_0
\]
Integration of Eq. \ref{eq:rho_evolution} yields:
\begin{equation}
\rho(t) = \rho(0) \exp\left[ -2 \nu_0 N_{i,\infty} t - \frac{\nu_0 (N_{i,0} - N_{i,\infty})}{\lambda_1} \left( 1 - e^{-\lambda_1 t} \right) \right]
\end{equation}

\noindent
For large times, $t \gg t_{c1}$, this simplifies to:
\begin{equation}
\rho(t) \approx \rho(0) \exp\left[ -\frac{\nu_0 (N_{i,0} - N_{i,\infty})}{\lambda_1} \right] \exp\left[ -2 \nu_0 N_{i,\infty} t \right]
\end{equation}

Ignoring internal relaxations, the mean charge evolution equation simplifies to:
\begin{equation}
\frac{d\bar{q}}{dt} = -(\beta_0 N_{p} + \nu_0 N) \bar{q} + \beta_0 \rho \approx -(\beta_0 N_{p} + \nu_0 N_{i,\infty}) \bar{q} + \beta_0 \rho_0' e^{-2 \nu_0 N_{i,\infty} t},
\end{equation}
where
\[
\rho_0' = \rho_0 \exp\left( -\frac{\nu_0 (N_{i,0} - N_{i,\infty})}{\lambda_1} \right).
\]

\noindent
Assuming $\rho_0' \approx \rho_0$, the solution for $\bar{q}(t)$ becomes:
\begin{equation}
\bar{q}(t) = q_0 e^{-\lambda_2 t} + \frac{\beta_0 N_{p} q_0}{\lambda_2 - \lambda_0} \left( e^{-\lambda_0 t} - e^{-\lambda_2 t} \right),
\label{qt}
\end{equation}
\noindent
where
\begin{align}
\lambda_0 &= 2 \nu_0 N_{i,\infty}, \\
\lambda_1 &= 2 \alpha_0 N_{i,\infty} + \beta_0 N_{p} = 2 \sqrt{\alpha_0 S + \frac{1}{4} \beta_0^2 N_{p}^2}, \\
\lambda_2 &= \beta_0 N_{p} + \nu_0 N_{i,\infty} = \sqrt{\alpha_0 S} \left[ k \left( 2 - \frac{\nu_0}{\alpha_0} \right) + \sqrt{k^2 + 1} \right].
\end{align}

Equation~\ref{qt} is the required expression for charge relaxation. It indicates that at higher aerosol concentrations, there are two characteristic time constants, namely $\lambda_0$ and $\lambda_2$. At zero aerosol concentration, $\lambda_2 = \nu_0 N_{i,\infty} = \lambda_0 / 2$. However, as the aerosol concentration increases, both time constants begin to play a significant role.
Hence, the neutralization rate of a concentrated aerosol cannot be described by a single neutralization coefficient. One must explicitly use the expression in Eq.~\eqref{qt} to estimate the degree of neutralization in a given situation. The concept of the traditional “$N_it$ product” should therefore be abandoned for high-density aerosol systems.

\bibliography{Refs}
\end{document}